\documentclass[footinbib,aps,pra,twocolumn,superscriptaddress,groupedaddress,a4paper]{revtex4}  % Uncomment for review/submission version
\usepackage{graphicx}
\usepackage{dcolumn}
\usepackage{bm}
\usepackage{amssymb}
\usepackage{amsmath}
\usepackage{subfigure}
\usepackage{color}
\usepackage{lscape}
\usepackage{siunitx}
\graphicspath{ {./Figures/}}

\begin{document}

\widetext
\title{Electronic and optical properties of Si$_{x}$Ge$_{1-x-y}$Sn$_{y}$ alloys lattice-matched to Ge}

%%%%%%%%%%%%%%%%%%%%%%%%%%%%%%%%%%%%%%%%
%%%% Authors, affiliations and date %%%%
%%%%%%%%%%%%%%%%%%%%%%%%%%%%%%%%%%%%%%%%

\author{Phoebe M.~Pearce}
\email{p.pearce@unsw.edu.au} % Email address must be kept before affiliation(s)
\altaffiliation[Present address: ]{School of Photovoltaic and Renewable Energy Engineering, University of New South Wales,
Sydney, NSW, Australia} 
\affiliation{Department of Physics, Imperial College London, South Kensington, London SW7 2AZ, United Kingdom}

\author{Christopher A.~Broderick}
\email{c.broderick@umail.ucc.ie} % Email address must be kept before affiliation(s)
\affiliation{Tyndall National Institute, University College Cork, Lee Maltings, Dyke Parade, Cork T12 R5CP, Ireland}
\affiliation{Department of Physics, University College Cork, Cork T12 YN60, Ireland}

\author{Michael P.~Nielsen}
\affiliation{School of Photovoltaic and Renewable Engineering, University of New South Wales, Sydney, New South Wales 2052, Australia}

\author{Andrew D.~Johnson}
\affiliation{IQE plc., Pascal Close, St. Mellons, Cardiff CF3 0LW, United Kingdom}

\author{Nicholas J.~Ekins-Daukes}
\affiliation{Department of Physics, Imperial College London, South Kensington, London SW7 2AZ, United Kingdom}
\affiliation{School of Photovoltaic and Renewable Engineering, University of New South Wales, Sydney, New South Wales 2052, Australia}

\date{\today}

%%%%%%%%%%%%%%%%%%%%%%%%%%%%%%%%%%%%%%%%%%%%%
%%%% Abstract, keywords and PACS numbers %%%%
%%%%%%%%%%%%%%%%%%%%%%%%%%%%%%%%%%%%%%%%%%%%%

\begin{abstract}
We present a combined experimental and theoretical analysis of the evolution of the near-band gap electronic and optical properties of Si$_{x}$Ge$_{1-x-y}$Sn$_{y}$ alloys lattice-matched to Ge and GaAs substrates. We perform photoreflectance (PR) and photoluminescence (PL) measurements on Si$_{x}$Ge$_{1-x-y}$Sn$_{y}$ epitaxial layers grown via chemical vapour deposition, for Si (Sn) compositions up to $x = 9.6$\% ($y = 2.5$\%). Our measurements indicate the presence of an indirect fundamental band gap, with PL observed $\approx$ 200--250 meV lower in energy than the direct $E_{0}$ transition identified by PR measurements. The measured PL is Ge-like, suggesting that the alloy conduction band (CB) edge is primarily derived from the Ge L-point CB minimum. Interpretation of the PR and PL measurements is supported by atomistic electronic structure calculations. Effective alloy band structures calculated via density functional theory confirm the presence of an indirect fundamental band gap, and reveal the origin of the observed inhomogeneous broadening of the measured optical spectra as being alloy-induced band hybridisation occurring close in energy to the CB edge. To analyze the evolution of the band gap, semi-empirical tight-binding (TB) calculations are employed to enable calculations for large supercell sizes. TB calculations reveal that the alloy CB edge is hybridized in nature, consisting at low Si and Sn compositions of an admixture of Ge L-, $\Gamma$- and X-point CB edge states, and confirm that the alloy CB edge retains primarily Ge L-point CB edge character. Our experimental measurements and theoretical calculations confirm a direct transition energy close to 1 eV in magnitude for Si and Sn compositions $x = 6.8$ -- 9.6\% and $y = 1.6$ -- 2.2\%.
\end{abstract}

% \keywords{0.0X}
% \pacs{0.0X}

\maketitle

%%%%%%%%%%%%%%%%%%%%%%%%%%%%%%%%%
%%%% Section 1: Introduction %%%%
%%%%%%%%%%%%%%%%%%%%%%%%%%%%%%%%%

\section{Introduction}
\label{sec:introduction}

% General introduction: SiGeSn alloys and multi-junction solar cell applications

Si$_{x}$Ge$_{1-x-y}$Sn$_{y}$ alloys have attracted increasing research interest due to their potential for a variety of device applications \cite{Wirths2016,Moutanabbir_APL_2021}, including diode lasers \cite{Sun_JAP_2010,Zhou2019}, light-emitting diodes \cite{Beeler2013,VondenDriesch2017}, photovoltaics \cite{Fang2008,Ventura2015,Beeler2013,Roucka2016a}, photodiodes \cite{Beeler2012}, and tunneling field-effect transistors \cite{Wirths2013,Sant_IEEEJEDS_2015}. For applications in multi-junction solar cells, a material lattice-matched to Ge or GaAs having an absorption edge close to 1 eV is especially useful as it can be incorporated into well-developed device architectures \cite{King2009}. The composition range for Si$_{x}$Ge$_{1-x-y}$Sn$_{y}$ alloys suitable for application as a 1 eV junction in a multi-junction solar cell is Ge-rich ($\gtrsim 85$\%), with a Si to Sn composition ratio $x~:~y \approx 3.7~:~1$ required to achieve lattice matching. Beginning from Ge, incorporation of Si increases the energy of both the fundamental indirect band gap (L$_{6c}$-$\Gamma_{8v}$; 0.67 eV in Ge) and the lowest-energy direction transition ($\Gamma_{7c}$-$\Gamma_{8v}$; 0.80 eV in Ge), while simultaneously reducing the lattice constant due to the lower covalent radius of Si compared to Ge. Conversely, incorporation of Sn acts to reduce the magnitude of the indirect and direct band gaps, while simultaneously increasing the lattice constant \cite{Moontragoon_JAP_2012,Ventura2015,Wendav2016}. By balancing the relative composition of Si and Sn in the alloy, it is therefore possible to engineer a material having higher fundamental band gap and absorption edge energies than Ge, but the same lattice constant \cite{Aella2004,Soref2007,Moontragoon_JAP_2012,Ventura2015}.

% What we do in this paper

Here, we present a combined experimental and theoretical analysis of the evolution of the electronic and optical properties of Si$_{x}$Ge$_{1-x-y}$Sn$_{y}$ alloys lattice-matched to Ge and GaAs. Experimentally, we describe sample growth via low-pressure chemical vapour deposition (CVD), while the direct inter-band optical transitions are probed using photoreflectance (PR) measurements. Photoluminescence (PL) measurements are employed to investigate the fundamental indirect band gap. Theoretically, we analyze the electronic structure evolution with alloy composition firstly using first principles calculations based on density functional theory (DFT), which we apply to compute effective (unfolded) band structures for alloy special quasi-random structures (SQSs) \cite{Zunger_PRL_1990,Wei_PRB_1990}. These calculations identify strong alloy-induced band hybridisation close in energy to the conduction band (CB) edge in energy, which we then probe via large-supercell semi-empirical calculations based on a parametrized valence force field (VFF) potential and tight-binding (TB) Hamiltonian \cite{Halloran_OQE_2019}.

We focus on the evolution of the near-band electronic and optical properties, quantifying the nature of the indirect and direct inter-band transitions critical to optoelectronic and photovoltaic device applications. Our measurements and calculations demonstrate that the fundamental band gap in lattice-matched Si$_{x}$Ge$_{1-x-y}$Sn$_{y}$ alloys suitable for solar cell applications remains indirect, but is larger than the fundamental L$_{6c}$-$\Gamma_{8v}$ band gap of Ge. The lowest-energy direct transition, which dominates the absorption edge of the material, is shown to be readily tunable to the $\approx 1$ eV values required for optimized three- and four-junction solar cells. In a solar cell, the maximum achievable sub-cell current is limited by the fundamental band gap, while the direct absorption edge energy is important from the perspective of current-matching. In addition to providing insight into the complexities of the electronic and optical properties of this emerging material system, our results suggest the potential to achieve higher current and improved current-matching in multi-junction solar cells having a Si$_{x}$Ge$_{1-x-y}$Sn$_{y}$ 1 eV junction.

% Organisation of this paper

% The remainder of this paper is organized as follows. In Section~\ref{sec:methodology_experimental} we describe the experimental methods used: the growth and characterisation of the investigated Si$_{x}$Ge$_{1-x-y}$Sn$_{y}$ samples, and details of the PR and PL spectroscopic measurements and data analysis. In Section~\ref{sec:methodology_theoretical} we describe our small-supercell first principles, and large-supercell semi-empirical, atomistic electronic structure calculations and their analysis. The results of our measurements and calculations and presented and discussed in Section~\ref{sec:results}. Finally, in Section~\ref{sec:conclusions} we summarize and conclude.

%%%%%%%%%%%%%%%%%%%%%%%%%%%%%%%%%%%%%%%%%
%%%% Section 2: Experimental methods %%%%
%%%%%%%%%%%%%%%%%%%%%%%%%%%%%%%%%%%%%%%%%

\section{Experimental methods}
\label{sec:methodology_experimental}

% Section 2.1 - Sample growth and characterisation

\subsection{Sample growth and characterisation}
\label{sec:methodology_samples}

\begin{figure}[t!]
 \centering
 \includegraphics[width=0.45\textwidth]{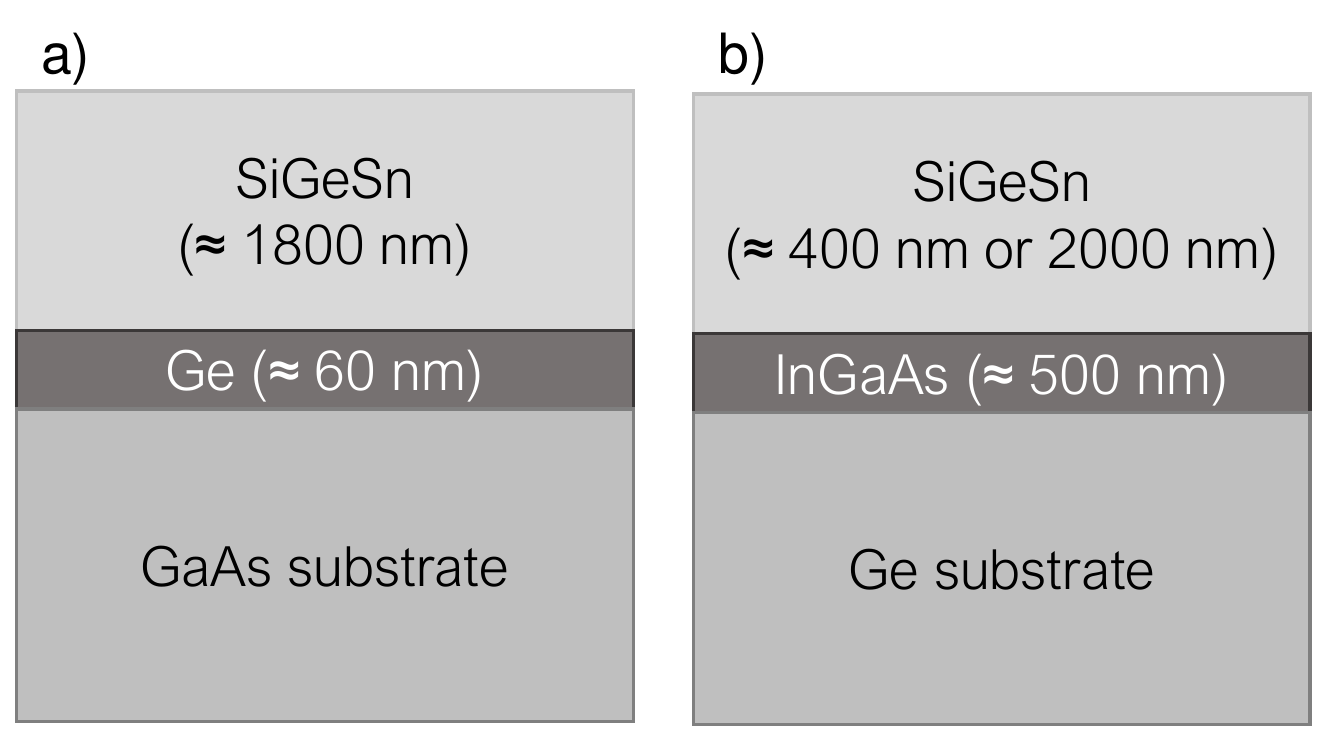}
 \caption{Schematic illustration of the structure of the epitaxial Si$_{x}$Ge$_{1-x-y}$Sn$_{y}$ samples investigated in this work. (a) Sample set A: grown on GaAs substrates, with a thin Ge seed layer. (b) Sample set B: grown on Ge substrates, with a thick lattice-matched In$_{0.012}$Ga$_{0.988}$As buffer layer. Layer thicknesses indicated in (a) and (b) are the nominal thicknesses targeted during CVD growth.}
 \label{fig:layerstructure}
\end{figure}

Si$_{x}$Ge$_{1-x-y}$Sn$_{y}$ samples were grown via low-pressure CVD using an ASM Epsilon 2000 reactor, with germane (GeH$_{4}$), disilane (Si$_{2}$H$_{6}$) and tin chloride (SnCl$_{4}$) precursors, and using H$_{2}$ as the carrier gas. Two sets of samples were grown across the composition range of interest. The first, set A, were grown on GaAs substrates with a thin ($\approx 60$ nm) Ge seed layer. The second, set B, were grown on Ge substrates which were overgrown via metal-organic vapour phase epitaxy with a $\approx$ 500 nm buffer layer of In$_{0.01}$Ga$_{0.99}$As lattice-matched to Ge. The layer structures of sample sets A and B are depicted schematically in Figs.~\ref{fig:layerstructure}(a) and~\ref{fig:layerstructure}(b), respectively. Samples were grown for a range of Si and Sn compositions $x$ and $y$, aiming in all cases for a composition ratio $x:y \approx 3.7:1$ to achieve lattice matching to Ge, with this composition ratio derived under the assumption that the alloy lattice constant $a(x,y)$ obeys Vegard's law \cite{Vegard1921}, and using room-temperature Si, Ge and $\alpha$-Sn lattice constants \cite{Madelung2002}. Both sets of samples include three different combinations of Si and Sn composition, and for sample set B two different structures having thin ($\sim 0.4$ \si{\micro}m) and thick ($\sim 2$ \si{\micro}m) Si$_{x}$Ge$_{1-x-y}$Sn$_{y}$ layers with nominally equal alloy compositions were grown, yielding a total of nine samples with six different alloy compositions. The resulting sample structures, Si and Sn compositions, and sample layer thicknesses (obtained via fits to measured PR data) are summarized in Table~\ref{tab:samples}. The Si and Sn compositions of the samples were measured using scanning electron microscopy with energy-dispersive x-ray (SEM-EDX) spectroscopy. These composition measurements, in addition to x-ray diffraction and spectroscopic ellipsometry measurements for these samples, are discussed in Ref.~\cite{Pearce2021}. The sample layer thicknesses reported in Ref.~\cite{Pearce2021}, obtained via fitting to spectroscopic ellipsometry data, are in good agreement with the values obtained here via fitting to PR data (cf.~Section~\ref{sec:methodology_photoreflectance}).

% Figure 1

% Table 1

\begin{table}[ht!]
    \caption{\label{tab:samples} Details of the Si$_{x}$Ge$_{1-x-y}$Sn$_{y}$ samples investigated. Sample set A were grown on GaAs substrates, using a Ge buffer layer (cf.~Fig.~\ref{fig:layerstructure}(a)). Sample set B were grown on Ge substrates, using a In$_{0.012}$Ga$_{0.988}$As buffer layer (cf.~Fig.~\ref{fig:layerstructure}(b)). Listed Si$_{x}$Ge$_{1-x-y}$Sn$_{y}$ layer thicknesses $d$ were obtained from fits to measured PR spectra. Two sets of B samples were grown: ``thin'' Si$_{x}$Ge$_{1-x-y}$Sn$_{y}$ layers having thickness $\sim 0.4$ \si{\micro}m, and ``thick'' layers having thickness $\sim 2$ \si{\micro}m. Listed Si and Sn compositions $x$ and $y$ are the mean value and standard deviation obtained from ten SEM-EDX measurements performed on each sample. Compositions measurements for the Set B samples were performed only for the thicker samples, with the thinner samples grown under the same conditions and assumed to have the same composition.}
    \begin{ruledtabular}
        \begin{tabular}{ccccc}
            Sample      & $d$ (nm)  & Si, $x$ (\%)  & Sn, $y$ (\%)  & Ratio, $x : y$ \\
            \hline
            A1          & 1811      & 4.6 $\pm$ 0.2       & 0.9 $\pm$ 0.1       & 5.11 : 1 \\
            A2          & 1746      & 6.8 $\pm$ 0.4       & 1.6 $\pm$ 0.1       & 4.25 : 1 \\
            A3          & 1828      & 9.6 $\pm$ 0.4       & 2.2 $\pm$ 0.1       & 4.36 : 1 \\
            \hline
            B1 (thin)   & 332       &                     &                     &          \\
            B2 (thin)   & 362       &                     &                     &          \\
            B3 (thin)   & 353       &                     &                     &          \\
            \hline
            B1 (thick)  & 2073      & 2.6 $\pm$ 0.3       & 0.8 $\pm$ 0.2       & 3.25 : 1 \\
            B2 (thick)  & 2017      & 5.9 $\pm$ 0.2       & 1.5 $\pm$ 0.1       & 3.93 : 1 \\
            B3 (thick)  & 2036      & 8.4 $\pm$ 0.2       & 2.5 $\pm$ 0.1       & 3.36 : 1 \\
        \end{tabular}
    \end{ruledtabular}
\end{table}

% Section 2.2 - Photoreflectance spectroscopy

\subsection{Photoreflectance spectroscopy}
\label{sec:methodology_photoreflectance}

The energy $E_{0}$ associated with the direct (optical, $\Delta \textbf{k} = 0$) transition between the zone-center $\Gamma_{8v}$ valence band (VB) and $\Gamma_{7c}$ CB edges, and the energy $E_{0} + \Delta_{0}$ associated with the direct transition between the zone-center $\Gamma_{7c}$ CB edge and the $\Gamma_{7v}$ spin-split-off (SO) VB, were studied using PR. PR is a contactless modulation spectroscopy technique that allows measurement of transition energies associated with direct inter-band transitions in semiconductors via modulation of the internal electric field as carriers are photogenerated with a laser pump beam. The Si$_{x}$Ge$_{1-x-y}$Sn$_{y}$ samples were doped (p-type, $n_a \approx 5 \times 10^{15}$ cm$^{-3}$). This generates an internal electric field due to pinning of the Fermi energy at the sample surface, which will have an excess of majority carriers (holes). When carriers are generated at the sample surface by the pump beam, this internal field causes the electrons and holes to be separated with the minority carriers (electrons) swept towards the surface. This carrier separation acts to screen the internal electric field, and thus the field is modulated by the laser \cite{Patane_SSMS_2012}. Sample reflectivity is measured via a probe beam, both with and without the effect of the pump beam, allowing the change in reflectivity $\Delta R$ due to the pump beam to be determined. PR is therefore a differential measurement technique, making it highly sensitive to critical points in the material band structure. In order to eliminate the effect of PL due to the laser pump beam from the PR measurements, and measure $R$ and $\Delta R$ simultaneously, both the probe and pump beams were chopped at different frequencies in a double demodulation method \cite{Amirtharaj1995}. As such, the $\Delta R$ signal was measured at the sum of the frequencies of the probe and pump beams, while the $R$ signal was directly measured at the probe chopping frequency. $\Delta R$ and $R$ can thus be measured using a single detector and extracted via lock-in detection, which in our measurements was performed using a Zurich Instruments HF2 lock-in amplifier, with signal gain accomplished using a Zurich Instruments HF2TA current amplifier. The pump beam used is a GEM 532 nm (2.33 eV) continuous-wave laser incident at an angle of $\approx 5^\circ$ to the sample surface with an intensity of 164 W cm$^{-2}$. The probe beam was generated using a Bentham IL1 100 W halogen lamp coupled to a Bentham TMC300 monochromator. The spectral range is limited by the laser wavelength (532 nm) at short wavelength (since there will be significant noise around the laser wavelength), and by the detector response at long wavelength (1700 nm). The probe beam was focused on the sample at an angle of 45$^\circ$, and the reflected beam was measured using a Thorlabs DET10N2 biased InGaAs detector with a detection range of 500--1700 nm.

Critical points in the measured PR spectra were analyzed by fitting the experimental data to an expression formed of a sum of third-derivative functional forms \cite{Aspnes1973}

\begin{eqnarray}
    \frac{ \Delta R }{ R } &=& \text{Re} \left\{ \frac{ A_{0} \, e^{ i \Phi_{0} } }{ \left( E - E_{0} + i \Gamma_{0} \right)^{5/2} } + \frac{ A_{\text{ex}} \, e^{ i \Phi_{\text{ex}} } }{ \left( E - E_{\text{ex}} + i \Gamma_{\text{ex}} \right)^{2} } \right. \nonumber \\
    &+& \left. \frac{ A_{\Delta_{0}} \, e^{ i \Phi_{\Delta_{0}} } }{ \left( E - ( E_{0} + \Delta_{0} ) + i \Gamma_{\Delta_{0}} \right)^{5/2} } \right\} \, , \label{eq:3DPR}
\end{eqnarray}

\noindent
where the energetic (lifetime) broadening $\Gamma$, amplitude $A$ and phase $\Phi$ in each term are assumed to be independent of photon energy $E$.

The first and third terms in Eq.~\eqref{eq:3DPR} represent the third derivative with respect to energy of the expected form of the dielectric function around a three-dimensional $M_{0}$-type critical point \cite{Adachi1988,DCosta2006b}, and correspond respectively to contributions to $\frac{ \Delta R }{ R }$ from CB-VB ($\Gamma_{7c}$-$\Gamma_{8v}$) and CB-SO ($\Gamma_{7c}$-$\Gamma_{7v}$) transitions having respective energies $E_{0}$ and $E_{0} + \Delta_{0}$. The second term represents the excitonic contribution to the third derivative of the dielectric function close in energy to the $E_{0}$ critical point, at energy $E_{\text{ex}}$. The measured PR spectrum for a Ge (substrate) reference sample shows a symmetrical peak shape close in energy to the direct $\Gamma_{7c}$-$\Gamma_{8v}$ transition \cite{Adachi1988} having three sharp extrema. It is not possible to achieve this peak shape by fitting only a single critical point feature at the $E_{0}$ feature, hence the inclusion of the excitonic contribution. To analyze the PR data of the Si$_{x}$Ge$_{1-x-y}$Sn$_{y}$ samples, it was necessary to account for the impact of the clear thin-film interference effects observed in the PR data. The PR data for the Si$_{x}$Ge$_{1-x-y}$Sn$_{y}$ samples with layer thickness $\approx 2$ \si{\micro}m showed multiple oscillation fringes about, and below, the expected $E_{0}$ energy, while the data for the thinner set B samples shows more clearly defined $E_{0}$ and $E_{0} + \Delta_{0}$ features which blueshift with increasing Si and Sn composition.

Although fringes due to Franz-Keldysh oscillations can occur in PR measurements \cite{Shen2011}, these occur at energies above the critical point, while the oscillations observed here are below the $E_{0}$ energies expected from theory and observed in spectroscopic ellipsometry analysis of the same samples \cite{Pearce2021}. Comparing the data for the thicker and thinner set B samples also shows that these oscillations are primarily visible at energies below and close to the $E_{0}$ critical point feature. The width of the fringes corresponds to the width expected due to thin-film oscillations for light incident at 45$^\circ$ (the angle of incidence of the probe beam with respect to the sample surface). The measured PR spectra for the Si$_{x}$Ge$_{1-x-y}$Sn$_{y}$ samples were thus fitted using a novel combined critical point and interference fringe model \cite{supmat}, with the exception of the Ge reference sample, which was fitted using Eq.~\eqref{eq:3DPR} directly. Previous work investigating the appearance of interference fringes in PR spectra \cite{Huang1989,Kallergi1990,Ghosh1997} has attributed the fringes to either modulation of the refractive index of doped materials \cite{Huang1989,Kallergi1990}, or temperature changes in the sample \cite{Ghosh1997}. Our fitting approach assumes that the fringes are due to the former, and expands upon previous models by self-consistently computing the unmodulated and modulated optical constants from the critical point energies, and feeding these optical constants into a transfer-matrix method simulation \cite{Alonso-Alvarez2018} to compute the modulated and unmodulated reflectivity and yield the total PR spectrum. Full details of the fitting procedure are described in the Supplemental Material \cite{supmat}.

% Section 2.3 - Photoluminescence spectroscopy

\subsection{Photoluminescence spectroscopy}
\label{sec:methodology_photoluminescence}

PL measurements were performed using excitation from a continuous-wave Spectra-Physics Millennia V Nd:YVO$_{4}$ laser with a wavelength of 532 nm (2.33 eV), and an incident power between 100 mW and 3 W (corresponding to incident intensities of 0.12--3.68 kW cm$^{-2}$, taking into account the laser spot size). The high photon energy means that most of the incident photons are absorbed within a few hundred nm of the sample surface, which ensures carriers are excited in the Si$_{x}$Ge$_{1-x-y}$Sn$_{y}$ epitaxial layer rather than the Ge or GaAs substrate (assuming that the absorption in the Si$_{x}$Ge$_{1-x-y}$Sn$_{y}$ epitaxial layers considered here is Ge-like at high incident photon energies of 2.33 eV, $> 99$\% of the incident laser power will be absorbed in the first 500 nm of material). The resulting PL signal was dispersed through a 0.5 m Princeton Acton SP2500i spectrometer and detected using a liquid nitrogen cooled Hamamatsu G7754-01 extended InGaAs detector. The monochromator was scanned across a wavelength range of 1000--2100 nm. The PL signal was processed via standard phase sensitive detection techniques, using a Stanford SR830 lock-in amplifier. All PL measurements were performed at room temperature ($T = 293$ K).

Power-dependent PL measurements were performed for the Ge reference and Si$_{x}$Ge$_{1-x-y}$Sn$_{y}$ alloy samples, with the range of incident laser intensities used for each sample chosen depending on its observed luminescence intensity. The Ge substrate was found to have significantly higher PL intensity than any of the Si$_{x}$Ge$_{1-x-y}$Sn$_{y}$ samples, being approximately an order of magnitude higher in terms of both peak and integrated intensity compared to the Si$_{x}$Ge$_{1-x-y}$Sn$_{y}$ sample displaying the highest PL intensity, sample A1. Due to the observed large spread in luminescence intensities the PL measurements were not performed at the same incident intensities for all samples, but rather the samples were divided into two groups. The Ge reference sample, as well as the alloy samples A1 and A3 -- which had the highest intensities of the Si$_{x}$Ge$_{1-x-y}$Sn$_{y}$ samples -- were measured at intensities between 0.12 and 2.46 kW cm$^{-2}$. PL measurements for the four remaining samples -- A2, B1, B2 and B3 -- were performed at intensities between 1.23 and 3.68 kW cm$^{-2}$. Due to the high incident laser power, especially for the second set of samples, local sample heating in the illuminated spot may be significant. To analyze this effect we have, at a given incident laser power, (i) calculated the increase in lattice temperature via numerical solution of the heat equation, and (ii) estimated the carrier temperature by fitting a Boltzmann distribution to the high-energy tail of the measured PL spectrum \cite{supmat}. Firstly, (i) reveals that the maximum local increase in sample lattice temperature -- relative to the ambient room temperature $T = 293$ K -- is expected to be up to $\approx 70$ K ($\approx$ 100 K) for an incident intensity of 2.46 kW cm$^{-2}$ (3.68 kW cm$^{-2}$). Secondly, (ii) produces estimated carrier temperatures that are in close quantitative agreement with our calculated lattice temperatures \cite{supmat}, suggesting the presence of carrier populations in thermal equilibrium with the lattice. To minimize the effects of sample heating on the interpretation of our measured PL spectra, band gaps were obtained solely from fits to our lowest-power measurements.

Both the Ge and Si$_{x}$Ge$_{1-x-y}$Sn$_{y}$ samples show multiple clear peaks in their PL spectra. For Ge the distinct PL peaks can be attributed to indirect (phonon-assisted) radiative recombination pathways \cite{Lieten2012}, with the exception of the highest-energy peak close to 0.78 eV $\approx E_{0}$ which is attributed to direct ($\Delta \textbf{k} = 0$) radiative recombination across the $\Gamma_{7c}$-$\Gamma_{8v}$ transition. To quantify the relative strength of the observed multiple peaks in the measured PL spectra, each spectrum was fitted using a combination of Gaussian and Lorentzian line shapes. The specific line shape used to fit to each peak was chosen to provide the best fit to the measured line shape of that peak. Generally, it was found that most peaks were best described using a Gaussian line shape, with the exception of the highest-energy peak corresponding to emission across the direct transition, which were best described using a Lorentzian line shape. For each PL spectrum the number of distinct peaks to include in the fit was determined via through a combination of counting the visible peaks and fitting the data to identify the number of peaks. The fitting parameters employed for each peak were the center (transition) energy, linewidth broadening, and intensity. Full details of the fitting procedure are described in the Supplemental Material \cite{supmat}.

%%%%%%%%%%%%%%%%%%%%%%%%%%%%%%%%%%%%%%%%
%%%% Section 3: Theoretical methods %%%%
%%%%%%%%%%%%%%%%%%%%%%%%%%%%%%%%%%%%%%%%

\section{Theoretical methods}
\label{sec:methodology_theoretical}

% Section 3.1 - First principles: density functional theory

\subsection{First principles: density functional theory}
\label{sec:methodology_dft}

% Description of DFT calculations and determination of lattice-matched alloy compositions

Our alloy supercell DFT calculations for ternary Si$_{x}$Ge$_{1-x-y}$Sn$_{y}$ alloys are based closely upon those we have recently established for binary Ge$_{1-y}$Sn$_{y}$ alloys, details of which can be found in Ref.~\cite{Halloran_OQE_2019}. Calculation of structural properties (lattice relaxation) proceeds in the local density approximation (LDA), with subsequent electronic structure calculations performed using the Tran-Blaha modified Becke-Johnson (TB-mBJ) exchange-correlation potential \cite{Tran_PRL_2009}. Our DFT calculations were performed using the projector-augmented wave method \cite{Blochl_PRB_1994,Kresse_PRB_1999}, as implemented in the Vienna Ab-initio Simulation Package (VASP) \cite{Kresse_CMS_1996,Kresse_PRB_1996}. All calculations explicitly include spin-orbit coupling. Cut-off energies, \textbf{k}-point grids, choice of pseudopotentials, and convergence criteria for lattice relaxation are as in Ref.~\cite{Halloran_OQE_2019}. To determine combinations of Si and Sn composition $x$ and $y$ which maintain lattice matching to a Ge substrate, we estimate the alloy lattice constant $a( x, y )$ via linear interpolation (Vegard's law) between the LDA-calculated lattice constants $a = 5.403$, 5.647 and 6.480 \AA~for Si, Ge and $\alpha$-Sn, respectively \cite{Tanner_VFF_2021}. Setting $a ( x, y ) = a ( \text{Ge} )$ we compute a composition ratio $x : y = 3.41 : 1$ for lattice-matched Si$_{x}$Ge$_{1-x-y}$Sn$_{y}$. This composition ratio is unchanged if lattice constants calculated via hybrid functional (HSEsol) DFT are employed \cite{Tanner_VFF_2021}. We note that the lattice-matched composition ratio $x : y = 3.41 : 1$ obtained from our zero-temperature DFT calculations differs slightly from the ratio 3.7:1 estimated in Section~\ref{sec:methodology_samples} using measured lattice constants. We emphasize that the experimental lattice constants cited in Section~\ref{sec:methodology_samples} were measured at room temperature, and therefore expect that the difference in thermal expansion coefficient between Ge, Si and $\alpha$-Sn accounts for the difference in lattice-matched composition ratios. In our DFT (TB) calculations we employ the LDA-calculated composition ratio $x$:$y = 3.41$:1 for the generation of lattice-matched supercells, for consistency with our LDA-DFT (VFF) relaxation of these structures.

% Figure 2

\begin{figure*}[t!]
	\includegraphics[width=1.00\textwidth]{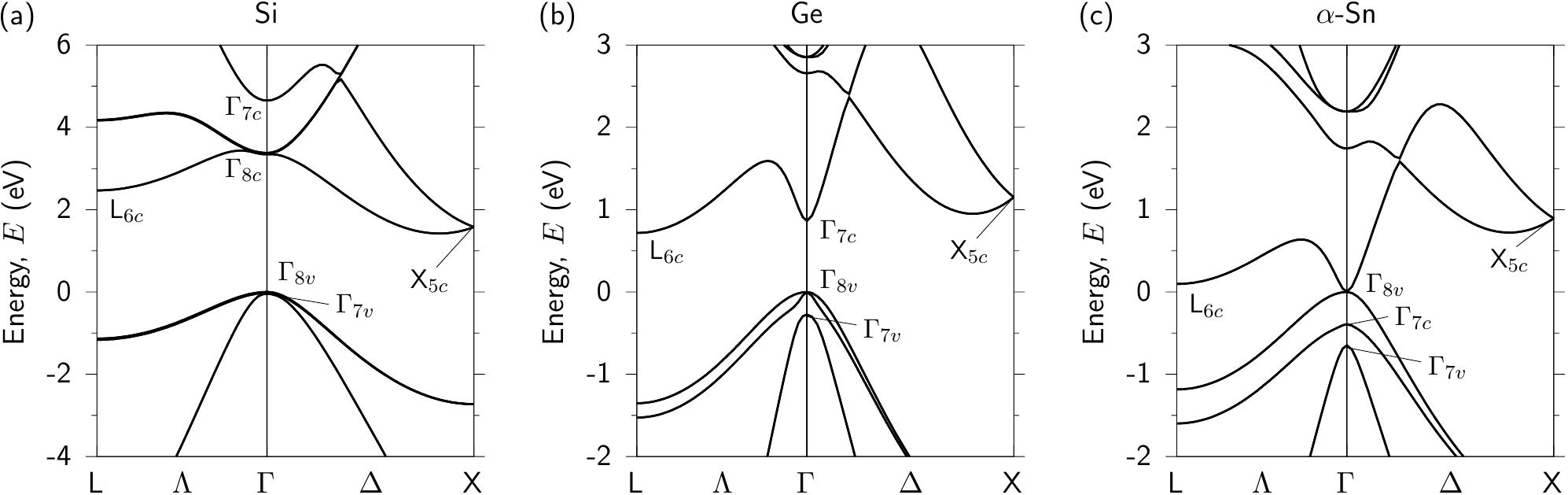}
	\caption{Band structures of (a) Si, (b) Ge, and (c) $\alpha$-Sn, calculated via DFT using the TB-mBJ exchange-correlation potential, and including spin-orbit coupling. \cite{Halloran_OQE_2019} In each case the zero of energy is chosen to lie at the $\Gamma_{8v}$ VB edge.}
	\label{fig:bulk_band_structures}
\end{figure*}

% Band structures of Si, Ge and Sn

Figure~\ref{fig:bulk_band_structures} shows the TB-mBJ DFT-calculated bulk band structures of (a) Si, (b) Ge, and (c) $\alpha$-Sn. Si and Ge are indirect-gap semiconductors, while $\alpha$-Sn is an inverted-gap semimetal possessing topologically non-trivial band ordering (in which the $s$-like $\Gamma_{7c}$ states lie intermediate in energy between the $p$-like $\Gamma_{8v}$ and $\Gamma_{7v}$ states) \cite{Booth_PR_1968,Groves_JPCM_1970}. We note that the increasing X-L-$\Gamma$ energy ordering of the high-symmetry point CB edge energies in Si is distinct from the L-$\Gamma$-X ordering in Ge. This, as well as the known tendency of Sn incorporation to induce $\Gamma$-L-X ordering -- i.e.~a direct fundamental band gap -- when incorporated in low concentrations in Ge, suggests that ternary Si$_{x}$Ge$_{1-x-y}$Sn$_{y}$ alloys can admit several distinct CB orderings \cite{Moontragoon_SST_2007}. It is therefore critical to understand the evolution of L-, $\Gamma$- and X-point CB edge states with composition in semiconducting Si$_{x}$Ge$_{1-x-y}$Sn$_{y}$ -- i.e.~at low Sn composition $y$ -- in order to develop a quantitative understanding of the alloy electronic structure.

% Choice of alloy supercells

Our previous analysis of Ge$_{1-y}$Sn$_{y}$ alloys revealed that Sn incorporation drives strong alloy-induced mixing (hybridisation) between Ge $\Gamma$- and L-point CB edge states \cite{Halloran_OQE_2019,Donnell_JPDAP_2021,Moutanabbir_APL_2021}. Alloy supercell calculations reveal that the Ge$_{1-y}$Sn$_{y}$ alloy CB minimum is formed of an admixture of Ge $\Gamma_{7c}$ and L$_{6c}$ states, with the nature of the alloy band gap evolving continuously from indirect (predominantly Ge L$_{6c}$-like) to direct (predominantly Ge $\Gamma_{7c}$-like) as the Sn composition $y$ increases \cite{Halloran_OQE_2019,Eales_SR_2019}. Since for photovoltaic applications we are interested in Ge-rich Si$_{x}$Ge$_{1-x-y}$Sn$_{y}$ alloys having $x + y \lesssim 15$\% -- i.e.~containing $\gtrsim 85$\% Ge -- we expect that alloy band mixing effects in the CB will also play an important role in determining the details of the Si$_{x}$Ge$_{1-x-y}$Sn$_{y}$ alloy electronic structure. As such, the choice of alloy supercells employed in our DFT calculations must be such that band mixing between $\Gamma$-, L- and X-point states is permitted to occur  \cite{Halloran_OQE_2019,Eales_SR_2019}. This is achieved by selecting 128-atom ($4 \times 4 \times 4$ face-centered cubic) supercells, in which the L and X points of the underlying diamond Brillouin zone fold back to the supercell zone center (at supercell wave vector $\textbf{K} = 0$). To provide a reliable description of short-range alloy disorder, we employ 128-atom SQSs in our DFT calculations: beginning with a Ge$_{128}$ supercell we substitute a given number of Ge atoms by Sn, determining the Sn composition $y$ and hence the Si composition $x = 3.41 \, y$ required to achieve lattice matching to Ge. The resulting alloy supercell is then relaxed using the LDA exchange-correlation functional, and the electronic structure of the relaxed supercell is computed using the TB-mBJ exchange-correlation potential. The SQSs used in our calculations were generated using the Alloy Theoretic Automated Toolkit (ATAT)  \cite{vandeWalle_JC_2002,vandeWalle_JC_2009}, by optimising the lattice correlation functions up to third-nearest neighbor distance about a given lattice site with respect to target correlation functions corresponding to a randomly disordered ternary alloy \cite{vandeWalle_JC_2013}.

To analyze the evolution of the Si$_{x}$Ge$_{1-x-y}$Sn$_{y}$ band gap, we compute the effective band structure \cite{Popescu_PRL_2010,Popescu_PRB_2012} by unfolding the TB-mBJ-calculated alloy SQS band structure onto a diamond primitive unit cell Brillouin zone defined by the LDA-relaxed alloy lattice constant $a ( x, y ) \approx a( \text{Ge} )$. This is achieved by calculating the total spectral weight $N ( \textbf{k}, E )$, where $\textbf{k}$ is the wave vector in the primitive unit cell Brillouin zone onto which the supercell band structure is being unfolded. $N ( \textbf{k}, E )$ at a given primitive cell wave vector \textbf{k}, in an energy interval of width $\Delta E$ centered at energy $E$, is computed by accumulating the individual supercell eigenstate spectral weights $W_{n\scalebox{0.7}{\textbf{K}}}$ within that energy interval as \cite{Medeiros_PRB_2014,Medeiros_PRB_2015}

\begin{equation}
      N( \textbf{k}, E ) = \int_{E - \frac{\Delta E}{2}}^{E + \frac{\Delta E}{2}} \sum_{n,\scalebox{0.7}{\textbf{K}}} W_{n\scalebox{0.7}{\textbf{K}}} \, \delta \left( E' - E_{n} ( \textbf{K} ) \right) \, \textrm{d}E' \, ,
      \label{eq:total_spectral_weight}
\end{equation}
 
\noindent
where $W_{n\scalebox{0.7}{\textbf{K}}} = \langle n \textbf{K} \vert \widehat{P} ( \textbf{K} \to \textbf{k} ) \vert n \textbf{K} \rangle$ is the spectral weight associated with the unfolding of supercell eigenstate $\vert n \textbf{K} \rangle$ from supercell wave vector \textbf{K} to primitive cell wave vector \textbf{k}, as described by the projection operator $\widehat{P} ( \textbf{K} \to \textbf{k} )$ \cite{Popescu_PRL_2010,Popescu_PRB_2012}. For our calculations here, we set $\Delta E = 5$ meV. All effective band structure calculations were performed using the BandUP code \cite{Medeiros_PRB_2014,Medeiros_PRB_2015}. We note that the integrand of Eq.~\eqref{eq:total_spectral_weight} is the spectral function $A ( \textbf{k}, E )$ \cite{Popescu_PRB_2012,Medeiros_PRB_2014}, so that the total spectral weight $N( \textbf{k}, E )$ is the integrated spectral function in the energy interval $[ E - \frac{ \Delta E }{ 2 }, E + \frac{ \Delta E }{ 2 } ]$.

% Section 3.2 - Semi-empirical: valence force field potential and tight-binding Hamiltonian

\subsection{Semi-empirical: valence force field potential and tight-binding Hamiltonian}
\label{sec:methodology_vff_and_tb}

% Choice of supercells for tight-binding electronic structure analysis

As described above, only states which fold to the same wave vector \textbf{K} in the supercell Brillouin zone are capable of hybridising and, as a consequence, careful choice of supercell lattice vectors is required to interrogate band hybridisation in small-supercell DFT calculations. More generally, the ability to analyze alloy-induced hybridisation effects in atomistic electronic structure calculations is limited by supercell size. As supercell size increases, more states fold to a given \textbf{K}, approaching the case of real alloy -- in which all states that can, by symmetry, hybridize will hybridize -- in the limit of ultra-large supercell size \cite{Zhang_PRB_2011,Moutanabbir_APL_2021}. Indeed, detailed analyzes of electronic structure evolution in conventional \cite{Zhang_PRB_2011} and highly-mismatched \cite{Usman_PRB_2011,Usman_PRA_2018} III-V alloys, as well as group-IV Ge$_{1-y}$Sn$_{y}$ \cite{Halloran_OQE_2019,Moutanabbir_APL_2021} and dilute carbide Ge$_{1-y}$C$_{y}$ alloys  \cite{Broderick_JAP_2019}, have demonstrated that obtaining quantitative accuracy in the computed electronic properties requires the use of supercells containing $\gtrsim 10^{3}$ atoms. While rigorous analysis of band hybridisation effects can be undertaken using DFT, the limitation of such calculations to supercell sizes $\sim 10^{2}$ atoms due to computational expense means that hybridisation can only occur between the limited number of states which fold to the same \textbf{K} point. As such, for alloys in which carrier localisation or band hybridisation play strong roles -- i.e.~where the widely-employed virtual crystal approximation (VCA) breaks down due to its omission of atomistic effects -- small-supercell DFT calculations qualitatively capture but often do not quantitatively predict properties observed in experimental measurements.

To overcome this limitation, we employ computationally less expensive semi-empirical TB calculations to enable high-throughput analysis of the electronic structure of large alloy supercells. Examining the band structures of Si, Ge and $\alpha$-Sn in Figs.~\ref{fig:bulk_band_structures}(a), \ref{fig:bulk_band_structures}(b) and~\ref{fig:bulk_band_structures}(c), we expect that the Si$_{x}$Ge$_{1-x-y}$Sn$_{y}$ alloy CB structure will evolve from having an indirect L-point minimum in Ge, towards having either an indirect $\Delta$-direction minimum (in Si-rich alloys), or an ``inverted'' direct band gap (in Sn-rich alloys) \cite{Moontragoon_SST_2007,Lan_PRB_2017,Polak_JPDAP_2017,Donnell_JPDAP_2021}. As such, to track the evolution of the alloy band gap we are interested primarily in analysing the evolution of character of alloy states lying close in energy to the CB minimum, and in allowing for hybridisation between L-, X- and $\Gamma$-point states. To achieve this we again select supercells in which both L- and X-point states fold back to the supercell zone center $\textbf{K} = 0$. We therefore perform calculations for 1024-atom ($8 \times 8 \times 8$ face-centered cubic) Si$_{M}$Ge$_{1024-M-N}$Sn$_{N}$ supercells, having respective Si and Sn compositions $x = \frac{M}{1024}$ and $y = \frac{N}{1024}$, with the numbers $M$ and $N$ of substitutional Si and Sn atoms chosen to remain as close to lattice-matched to Ge as possible ($M : N \approx 3.41 : 1$). The atomic positions in each alloy supercell are relaxed using a highly accurate (analytically parametrized) VFF potential \cite{Halloran_OQE_2019,Tanner_VFF_2021} -- implemented using the General Utility Lattice Program (GULP) \cite{Gale_JCSFT_1997,Gale_MS_2003,Gale_ZK_2005} -- and the alloy eigenstates at $\textbf{K} = 0$ are calculated via direct diagonalisation of the alloy supercell TB Hamiltonian \cite{Halloran_OQE_2019}. We note that our semi-empirical (VFF + TB) calculations here are based closely on those established and carefully benchmarked against DFT alloy supercell calculations for Ge$_{1-y}$Sn$_{y}$, full details of which can be found in Ref.~\cite{Halloran_OQE_2019}.

% Character of alloy conduction band states

To investigate the nature and evolution of the alloy CB structure, and to quantify alloy-induced hybridisation between Ge host matrix eigenstates, we take selected unperturbed Ge eigenstates $\vert n^{(0)} \rangle$, having energies $E_{n}^{(0)}$, and project them onto the complete spectrum of $\textbf{K} = 0$ Si$_{x}$Ge$_{1-x-y}$Sn$_{y}$ alloy supercell eigenstates $\vert m \rangle$, having energies $E_{m}$, as \cite{Usman_PRB_2011,Usman_PRA_2018}:

\begin{equation}
    G_{n} (E) = \sum_{m} \sum_{k = 1}^{ g ( E_{n}^{(0)} ) } f_{n,mk} \, T \left( E_{m} - E \right) \, ,
    \label{eq:character_spectrum}
\end{equation}

\noindent
where $g ( E_{n}^{(0)} )$ is the degeneracy of the Ge eigenstate $\vert n^{(0)} \rangle$, $T ( E_{m} - E )$ is a so-called ``top-hat'' filter having unit height for $E = E_{m}$, being otherwise zero, and $f_{n,mk} = \vert \langle n^{(0)} \vert m \rangle \vert^{2}$ is the squared projection of the alloy supercell eigenstate $\vert m \rangle$ onto the Ge reference eigenstates $\vert n^{(0)} \rangle$. By computing $G_{n} (E)$ for a series of Ge eigenstates it is possible for a given alloy supercell to explicitly identify the composition of each individual alloy eigenstate as a linear combination of the eigenstates of the Ge host matrix semiconductor. This allows direct interrogation of alloy band hybridisation effects and, undertaken as a function of alloy composition, enables the evolution of the alloy electronic structure to be quantified precisely \cite{Usman_PRB_2011,Usman_PRA_2018}. Equation~\eqref{eq:character_spectrum} defines a spectral function that is closely related to the $\textbf{K} = 0$ contribution to the spectral function $A ( \textbf{k}, E )$ employed in our DFT effective band structure calculations (cf.~Eq.~\eqref{eq:total_spectral_weight}). Here, $G_{n} (E)$ represents the contribution to the $\textbf{K} = 0$ Bloch character associated with an individual Ge host matrix eigenstate $\vert n^{(0)} \rangle$ which is, in general, spread over several alloy supercell eigenstates $\vert m \rangle$ due to alloy-induced hybridisation. Summing $G_{n} (E)$ over all Ge reference eigenstates $\vert n^{(0)} \rangle$ produces a spectral function representing the total Bloch character of all supercell eigenstates that back-fold to $\textbf{K} = 0$. In our TB analysis we compute and analyze $G_{n} (E)$ directly for individual Ge reference eigenstates, so as to explicitly quantify the contributions of individual Ge eigenstates to the hybridized Si$_{x}$Ge$_{1-x-y}$Sn$_{y}$ alloy eigenstates.

%%%%%%%%%%%%%%%%%%%%%%%%%%%%
%%%% Section 4: Results %%%%
%%%%%%%%%%%%%%%%%%%%%%%%%%%%

\section{Results}
\label{sec:results}

Interpretation of alloy electronic and optical properties is challenging due to breakdown of \textbf{k} selection associated with the reduction in translational symmetry arising from short-range alloy disorder and lattice relaxation. We therefore begin our discussion with the results of our small-supercell DFT and large-supercell TB electronic structure calculations in Secs.~\ref{sec:results_theory_dft} and \ref{sec:results_theory_tb}, in order to inform the analysis of our PR and PL measurements in Secs.~\ref{sec:results_experiment_pr} and \ref{sec:results_experiment_pl}. In Section~\ref{sec:results_theory_vs_experiment} we compare our theoretical calculations and experimental measurements of the composition-dependent direct and indirect band gaps.

% Figure 3

\begin{figure*}[t!]
	\includegraphics[width=1.00\textwidth]{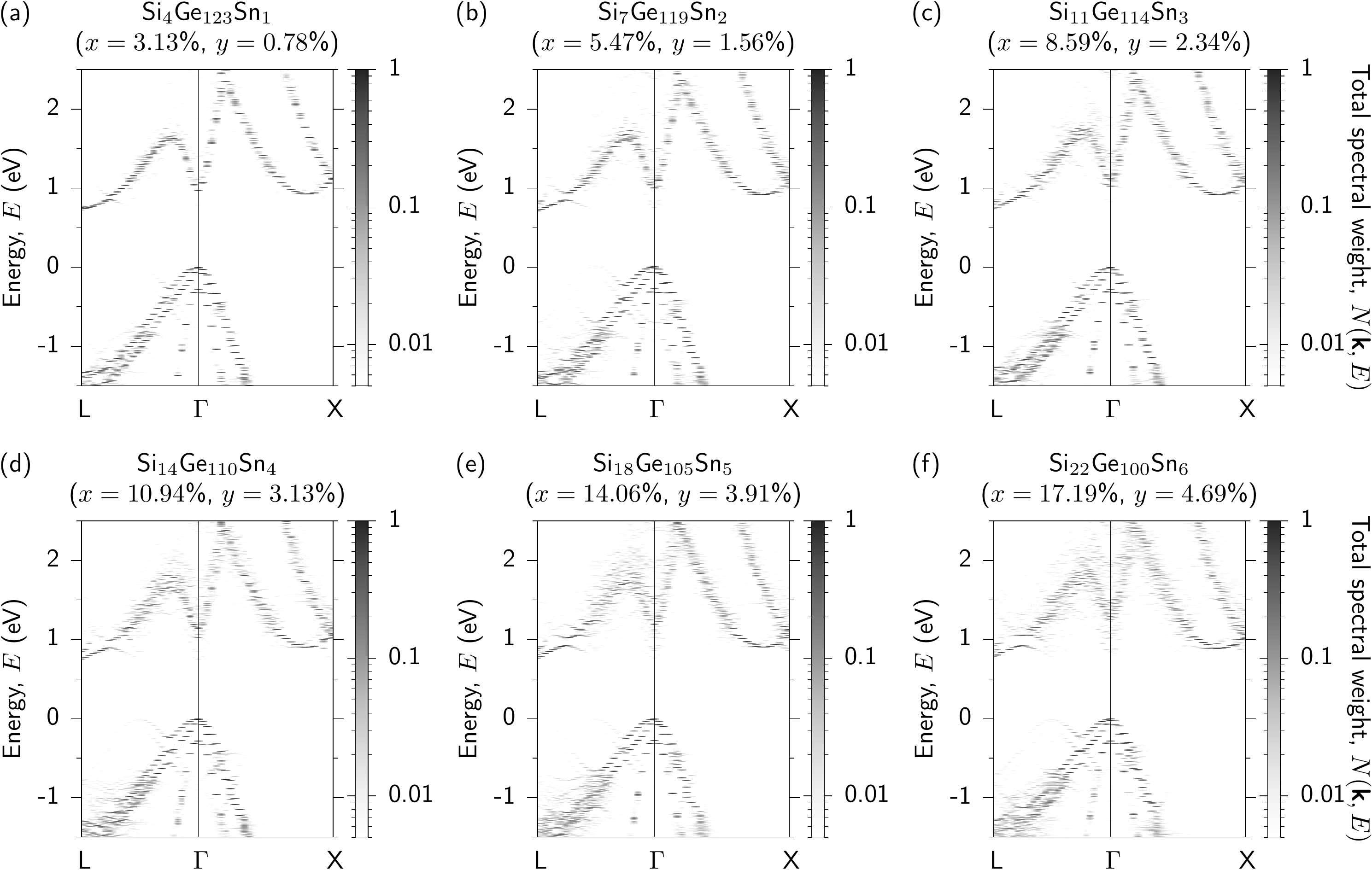}
	\caption{Effective (unfolded) alloy band structures, calculated via TB-mBJ DFT for relaxed 128-atom Si$_{x}$Ge$_{1-x-y}$Sn$_{y}$ alloy supercells. The supercells contain (a) 1, (b) 2, (c) 3, (d) 4, (e) 5, and (f) 6 Sn atoms, with the corresponding number of Si atoms chosen in each case to achieve lattice matching to Ge ($x$ : $y = 3.71$ : 1). The computed spectral weight $N$ is depicted on a logarithmic scale, to emphasize the distribution of Bloch character over a significant wave vector and energy ranges for states lying close in energy to the CB minimum. The zero of energy is chosen to lie at the alloy VB edge for each supercell.}
	\label{fig:unfolded_bands}
\end{figure*}

% Section 4.1 - Density functional theory: alloy effective band structures

\subsection{Density functional theory: alloy effective band structures}
\label{sec:results_theory_dft}

% Effective band structures: evolution of the alloy band gap

The results of our effective band structure calculations for lattice-matched Si$_{x}$Ge$_{1-x-y}$Sn$_{y}$/Ge 128-atom SQSs are summarized in Fig.~\ref{fig:unfolded_bands}. Here, the unfolded band structures are visualized in terms of their computed total Bloch character (spectral weight) $N( \textbf{k}, E )$, which we have plotted on a logarithmic scale in order to emphasize features indicative of, and resulting from, alloy band mixing effects. These calculations reveal several key qualitative features of the electronic structure of lattice-matched Si$_{x}$Ge$_{1-x-y}$Sn$_{y}$/Ge that are markedly different from the simplified description provided by the VCA.

The alloy CB structure is primarily Ge-like at low Si and Sn composition, with the CB minimum lying at the L-point, but rapidly evolves to become Si-like with increasing Si and Sn composition, with the CB minimum lying approximately 80\% of the way along the $\Delta$ direction between the $\Gamma$ and X points. Beginning with a Si$_{4}$Ge$_{123}$Sn$_{1}$ ($x = 3.13$\%, $y = 0.78$\%) SQS in Fig.~\ref{fig:unfolded_bands}(a) we note that the band structure is qualitatively similar to that of Ge, but that even at this low Si composition the CB minimum along the $\Delta$ direction is lower in energy than the $\Gamma$-point CB edge. This band ordering is already similar to that of pure Si, where the $\Gamma$-point CB edge lies higher in energy than the CB edge states at the L and X points. As the Si and Sn compositions increase in Figs.~\ref{fig:unfolded_bands}(b)--\ref{fig:unfolded_bands}(f) the magnitude of the calculated zero-temperature lowest-energy direct transition increases strongly, becoming equal to 1 eV by $x \approx 5.5$\% ($y \approx 1.6$\%; cf.~Fig.~\ref{fig:unfolded_bands}(b)). Simultaneously, the minimum of the lowest energy CB along the $\Delta$ direction is reduced in energy below the $\Gamma$-point CB edge. We note that our unfolded SQS band structures predict that lattice-matched Si$_{x}$Ge$_{1-x-y}$Sn$_{y}$/Ge alloys retain an indirect fundamental band gap, in accordance with previous model estimates \cite{Sun_JAP_2010} and small supercell empirical pseudopotential calculations \cite{Moontragoon_JAP_2012}. This behavior, which is in stark contrast to the emergence of a direct band gap in binary Ge$_{1-y}$Sn$_{y}$ alloys, is driven by the requirement to incorporate Si and Sn in the ratio 3.41:1 to maintain lattice matching to Ge or GaAs, with the impact of Si incorporation being to strongly reduce the energy of the $\Delta$ CB minimum relative to the zone-center CB edge \cite{Moontragoon_JAP_2012}. At the highest Si and Sn compositions shown in Fig.~\ref{fig:unfolded_bands} -- the Si$_{22}$Ge$_{100}$Sn$_{6}$ ($x = 17.19$\%, $y = 4.69$\%) SQS of Fig.~\ref{fig:unfolded_bands}(f) -- the zero-temperature band gap is $\approx 1.2$ eV in magnitude, while the L and $\Delta$ CB minima are approximately equal in energy. At higher Si compositions (not shown) the $\Delta$ minima reduces in energy below the L-point CB edge, with the increasing X-L-$\Gamma$ CB energy ordering then corresponding to that in pure Si for $x \gtrsim 20$\% ($y \gtrsim 6$\%).

% Effective band structures: alloy band mixing effects

Additionally, with increasing Si and Sn composition we note the emergence of significant energetic broadening of the CB Bloch character. This behavior is characteristic of alloy disorder and band mixing effects. The reduction of translational symmetry due to chemical inhomogeneity and lattice relaxation means that the wave vector \textbf{k} is no longer a ``good'' quantum number (i.e.~the crystal momentum is no longer a constant of carrier motion), with the resulting breakdown of \textbf{k}-selection broadening the bands at fixed energy $E$. Additionally, this symmetry breaking ensures that orbital components of primitive cell \textbf{k} states having equivalent symmetry which fold back to the same supercell wave vector \textbf{K} can hybridize with one another. In the $4 \times 4 \times 4$ face-centered cubic supercells considered here, the states which back-fold to $\textbf{K} = 0$ and can hence hybridize with $\Gamma$-point states are those having primitive cell lattice vectors \textbf{k} whose components are integer multiples of $\frac{ \pi }{ 2 a }$ \cite{Halloran_OQE_2019,Broderick_JAP_2019}. This includes L- and X-point states, so that our supercell calculations allow to track the band gap evolution in the presence of alloy-induced band mixing between direct and indirect CB edge states. Hybridisation between states which unfold from the same supercell wave vector \textbf{K} to different primitive wave vectors \textbf{k} are visible as ``echoes'' in $N( \textbf{k}, E )$ at equal $E$ and at the different \textbf{k} onto which the states unfold. This hybridisation, combined with the loss of strict \textbf{k}-selection, results in the distribution of Bloch character across the energies and wave vectors of the hybridising states. This further drives energetic broadening of the alloy CB Bloch character, which is then enhanced in the presence of short-range alloy disorder. For example, in Figs.~\ref{fig:unfolded_bands}(e) and~\ref{fig:unfolded_bands}(f) the emergence of a Bloch character ``tail'' at $\Gamma$ is visible at the same energies as the L- and X-point CB edge energies, reflecting that the loss of chemical and structural homogeneity drives hybridisation between these states which acts to spectrally broaden the zone-center alloy CB edge.

The presence of such features in our effective band structure calculations reveals that alloy band mixing effects are present in Si$_{x}$Ge$_{1-x-y}$Sn$_{y}$ alloys, being strongest close in energy to the CB edge, and hence that the direct alloy band gap is likely of hybridized character -- being primarily indirect, but possessing an admixture of direct character -- in the Ge-rich composition range under investigation \cite{Halloran_OQE_2019,Eales_SR_2019,Moutanabbir_APL_2021}. Given the small separation in energy between L-, $\Gamma$- and X-point CB edge states at these compositions, we expect that the hybridized zone-center alloy CB edge will be formed primarily of an admixture of the L$_{6c}$, $\Gamma_{7c}$ and X$_{5c}$ CB edge states of Ge (cf.~Section~\ref{sec:results_theory_tb}). In several of the calculated effective band structures we note that while an accumulation of $\Gamma$-point CB Bloch character in a given energy range enables a direct band gap to be defined in principle -- e.g.~by weighting the energy of individual alloy CB states by their computed Bloch character \cite{Donnell_JPDAP_2021} -- band mixing effects can lead to the emergence of $\Gamma$-point CB Bloch character at energies above and below the fundamental band gap. In this manner, alloy band mixing effects drive strong, inhomogeneous spectral broadening of the direct (optical, $\Gamma$-point) band gap. These qualitative features of the Si$_{x}$Ge$_{1-x-y}$Sn$_{y}$ band structure are markedly distinct from the simplified interpretation offered by the VCA (and related approaches that neglect atomistic effects), highlighting that direct atomistic calculations are required to provide quantitative understanding of the nature and evolution of the strongly perturbed CB structure.

Conversely, the band structure close in energy to the VB edge remains predominantly Ge-like across the range of compositions studied here. Si (Sn) incorporation in Ge acts to decrease (increase) the VB spin-orbit splitting energy $\Delta_{0}$, with these effects approximately cancelling one another when the Si to Sn composition ratio is chosen to maintain lattice matching to Ge, so that the alloy spin-orbit splitting energy remains approximately equal to that of Ge. Furthermore, compared to the calculated effective CB structure, we note minimal energetic broadening of the calculated Bloch character. This suggest that alloy band mixing effects have little impact on the band structure close in energy to the VB edge, similar to binary Ge$_{1-x}$Sn$_{x}$ alloys \cite{Halloran_OQE_2019,Donnell_JPDAP_2021}.

% DFT calculations: overall conclusions

Overall, our DFT calculations (i) predict that lattice-matched Si$_{x}$Ge$_{1-x-y}$Sn$_{y}$/Ge alloys retain an indirect fundamental band gap, and (ii) suggest that alloy band mixing effects, akin to those recently identified in Ge$_{1-y}$Sn$_{y}$ \cite{Halloran_OQE_2019,Eales_SR_2019,Moutanabbir_APL_2021}, play an important role in determining the nature and evolution of the alloy CB structure. The observed energetic broadening of the CB Bloch character in purely substitutional (defect-free) alloy supercell calculations suggests that the optical absorption and emission spectra of Si$_{x}$Ge$_{1-x-y}$Sn$_{y}$ are predicted to be characterized by strong inhomogeneous spectral broadening. This spectral broadening, identified in our calculations as being driven by alloy-induced CB hybridisation and enhanced by a loss of translational symmetry due to short-range alloy disorder, is expected to be an intrinsic property of Si$_{x}$Ge$_{1-x-y}$Sn$_{y}$ alloys. To obtain a quantitative understanding of the evolution of the electronic structure requires the treatment of large-scale alloy supercells, for which we turn to a semi-empirical approach.

% Section 4.2 - Tight-binding: evolution of conduction band states

\subsection{Tight-binding: evolution of alloy band edge states}
\label{sec:results_theory_tb}

% Figure 4

\begin{figure*}[t!]
	\includegraphics[width=1.00\textwidth]{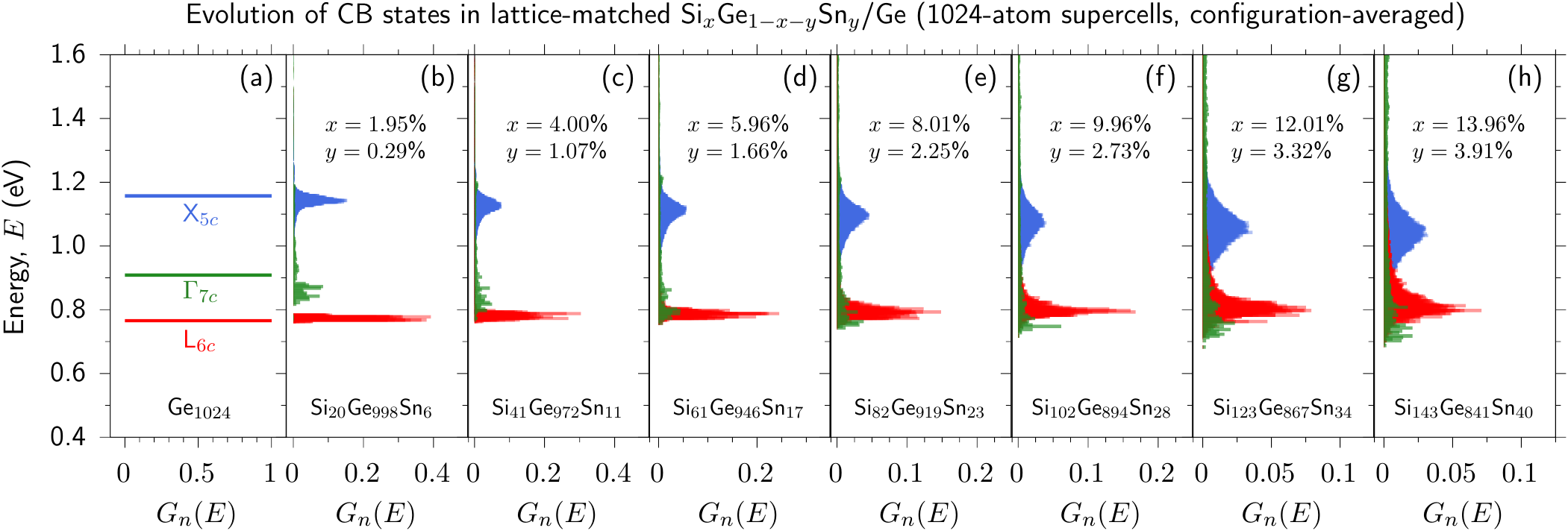}
	\caption{Calculated evolution of the configuration-averaged Ge L$_{6c}$ (red), $\Gamma_{7c}$ (green) and X$_{5c}$ (blue) character of the CB states in disordered, 1024-atom Si$_{x}$Ge$_{1-x-y}$Sn$_{y}$ alloy supercells lattice-matched to Ge ($x$ : $y = 3.71$ : 1). Panel (a) illustrates for reference that the L-, $\Gamma$- and X-point CB edge states in pure Ge respectively possess 100\% Ge L$_{6c}$, $\Gamma_{7c}$ and X$_{5c}$ character. Substitutional incorporation of Si and Sn -- panels (b)--(h) -- drives alloy-induced hybridisation, with the CB states in Ge-rich alloys formed of an admixture of Ge host matrix CB states. The reduction in symmetry associated with short-range alloy disorder results in inhomogeneous energetic broadening of the associated direct (green) and indirect (red and blue) CB edges. The zero of energy is taken to lie at the Ge VB edge.}
	\label{fig:cb_character}
\end{figure*}

Figure~\ref{fig:cb_character} shows the TB-calculated Ge L$_{6c}$ (red), $\Gamma_{7c}$ (green) and X$_{5c}$ (blue) character of the Si$_{x}$Ge$_{1-x-y}$Sn$_{y}$ alloy CB states as a function of composition for 1024-atom disordered alloy supercells. Here, the alloy CB L-, $\Gamma$- and X-point CB character is computed using Ge$_{1024}$ eigenstates $\vert n^{(0)} \rangle = \vert \text{L}_{6c}^{(0)} \rangle$, $\vert \Gamma_{7c}^{(0)} \rangle$ and $\vert \text{X}_{5c}^{(0)} \rangle$ in Eq.~\eqref{eq:character_spectrum}. For reference, Fig.~\ref{fig:cb_character}(a) shows the calculated Ge L$_{6c}$, $\Gamma_{7c}$ and X$_{5c}$ character in unperturbed Ge$_{1024}$. Here, the $\textbf{K} = 0$ eigenstates form a complete orthonormal basis set, with $\langle m^{(0)} \vert n^{(0)} \rangle = \delta_{nm}$, so that Eq.~\eqref{eq:character_spectrum} returns a value of unity at the energy $E_{n}^{(0)}$ of the reference Ge eigenstate $\vert n^{(0)} \rangle$. Results for alloy supercells in Figs.~\ref{fig:cb_character}(b) --~\ref{fig:cb_character}(h) are obtained via configurational averaging: at each alloy composition we construct, relax and compute the electronic structure of 25 distinct, randomly disordered supercells, and use the resulting alloy supercell eigenstates $\vert m \rangle$ in Eq.~\eqref{eq:character_spectrum}. Once the $G_{n} (E)$ spectra are computed for each of these 25 supercells, configurational averaging is performed by (i) accumulating the total Ge L$_{6c}$, $\Gamma_{7c}$ or X$_{5c}$ character calculated for each individual supercell into energy bins of width 5 meV, and (ii) averaging the total Ge L$_{6c}$, $\Gamma_{7c}$ or X$_{5c}$ character accumulated in each of these energy bins for alloy 25 distinct disordered supercells considered at that composition. The resulting $G_{n} (E)$ spectra of Figs.~\ref{fig:cb_character}(b) --~\ref{fig:cb_character}(h) provide detailed insight into the evolution of the alloy CB structure, effectively decomposing the calculated total Bloch character of the DFT unfolded alloy band structures of Fig.~\ref{fig:unfolded_bands} into its distinct L-, $\Gamma$- and X-point contributions.

Examining the evolution of the alloy CB structure described by Fig.~\ref{fig:cb_character}, we firstly note a blueshift and energetic broadening of the L-point CB edge Bloch character with increasing Si and Sn composition. Across the composition range investigated, the alloy CB edge retains primarily indirect Ge L$_{6c}$ character (red), but also acquires an admixture of direct Ge $\Gamma_{7c}$ character (green). By comparison, we note that the energetic broadening of the $\Gamma$-point Bloch character due to alloy band mixing is significantly enhanced, with Ge $\Gamma_{7c}$ character spread over an energy range extending upwards from the alloy CB edge by $\sim 1$ eV, even at low Si and Sn compositions $x = 4.00$\% and $y = 1.07$\% (cf.~Fig.~\ref{fig:cb_character}(c)). We note that this is primarily driven by hybridisation of $\Gamma$-point states with the large density of L-point states lying in the same energy range. Our calculations indicate that increasing the Si and Sn composition drives an admixture of Ge $\Gamma_{7c}$ character to the alloy CB edge, while the majority of the calculated Ge L$_{6c}$ character remains close in energy to the alloy CB edge, suggesting the presence of an indirect fundamental band gap. Finally, we note that the evolution of the X-point Bloch character is qualitatively similar to the L-point character, albeit redshifting in energy with increasing Si and Sn composition. The calculated Ge X$_{5c}$ character also undergoes inhomogeneous energetic broadening with increasing Si and Sn composition, although in a less pronounced manner than that computed for the $\Gamma_{7c}$ character.

As suggested by our DFT-calculated effective alloy band structures, our large-supercell TB calculations confirm that Si and Sn incorporation in Ge to form ternary Si$_{x}$Ge$_{1-x-y}$Sn$_{y}$ alloys primarily and strongly impacts the CB structure, with the VB structure remaining comparatively unperturbed (and, for lattice-matched compositions, quantitatively close to that of the Ge host matrix semiconductor). This was verified via additional TB calculations of the alloy VB maximum (Ge $\Gamma_{8v}$) and spin-split-off VB edge (Ge $\Gamma_{7v}$) character which, compared to the calculated CB character of Fig.~\ref{fig:cb_character}, were found to undergo minimal energy shifts or energetic broadening \cite{supmat}.

% Band gap evolution vs. experiment

Having tracked the evolution of the CB and VB edge states in lattice-matched Si$_{x}$Ge$_{1-x-y}$Sn$_{y}$ via atomistic electronic structure calculations, we turn our attention now to elucidating the implications of our calculations for the nature and evolution of the alloy band gap. The strong inhomogeneous energetic broadening of Ge CB character across a multiplicity of hybridized alloy CB states complicates interpretation of the alloy band gap, and hence comparison between theoretical calculations and experimental measurements. To facilitate comparison to experiment, we use the $G_{n} (E)$ spectra of Fig.~\ref{fig:cb_character} to compute the weighted average energy associated with the Ge L$_{6c}$-, $\Gamma_{7c}$- and X$_{5c}$-like alloy CB states and, in this manner, compute the composition-dependent band gaps that the associated indirect or direct optical transitions can be expected to track in experiment. We define this weighted average energy as

\begin{equation}
    \langle E_{n} \rangle = \sum_{m} \sum_{k = 1}^{ g ( E_{n}^{(0)} ) } f_{n,mk} \, E_{m} \, 
    \label{eq:weighted_average_energy}
\end{equation}

\noindent
where we use the Ge $\vert n^{(0)} \rangle$ character of each individual alloy supercell eigenstate $\vert m \rangle$ as the weight for the latter state's energy (cf.~Eq.~\eqref{eq:character_spectrum} and Fig.~\ref{fig:cb_character}). We note that Eq.~\eqref{eq:weighted_average_energy} is properly normalized due to the fact that the $\textbf{K} = 0$ eigenstates $\lbrace \vert n^{(0)} \rangle \rbrace$ and $\lbrace \vert m \rangle \rbrace$ of the Ge and Si$_{x}$Ge$_{1-x-y}$Sn$_{y}$ supercells form complete basis sets, so that the weights $f_{n,mk}$ sum to unity. As a consequence of the spectral theorem, Eq.~\eqref{eq:weighted_average_energy} is equivalent to the matrix element $\langle n^{(0)} \vert \widehat{H} (x,y) \vert n^{(0)} \rangle$ taken using the Ge host matrix supercell eigenstate $\vert n^{(0)} \rangle$, and the Hamiltonian $\widehat{H} (x,y)$ of the Si- and Sn-containing alloy supercell.

% Figure 5

\begin{figure*}[]
	\centering
	\includegraphics[width=0.98\textwidth]{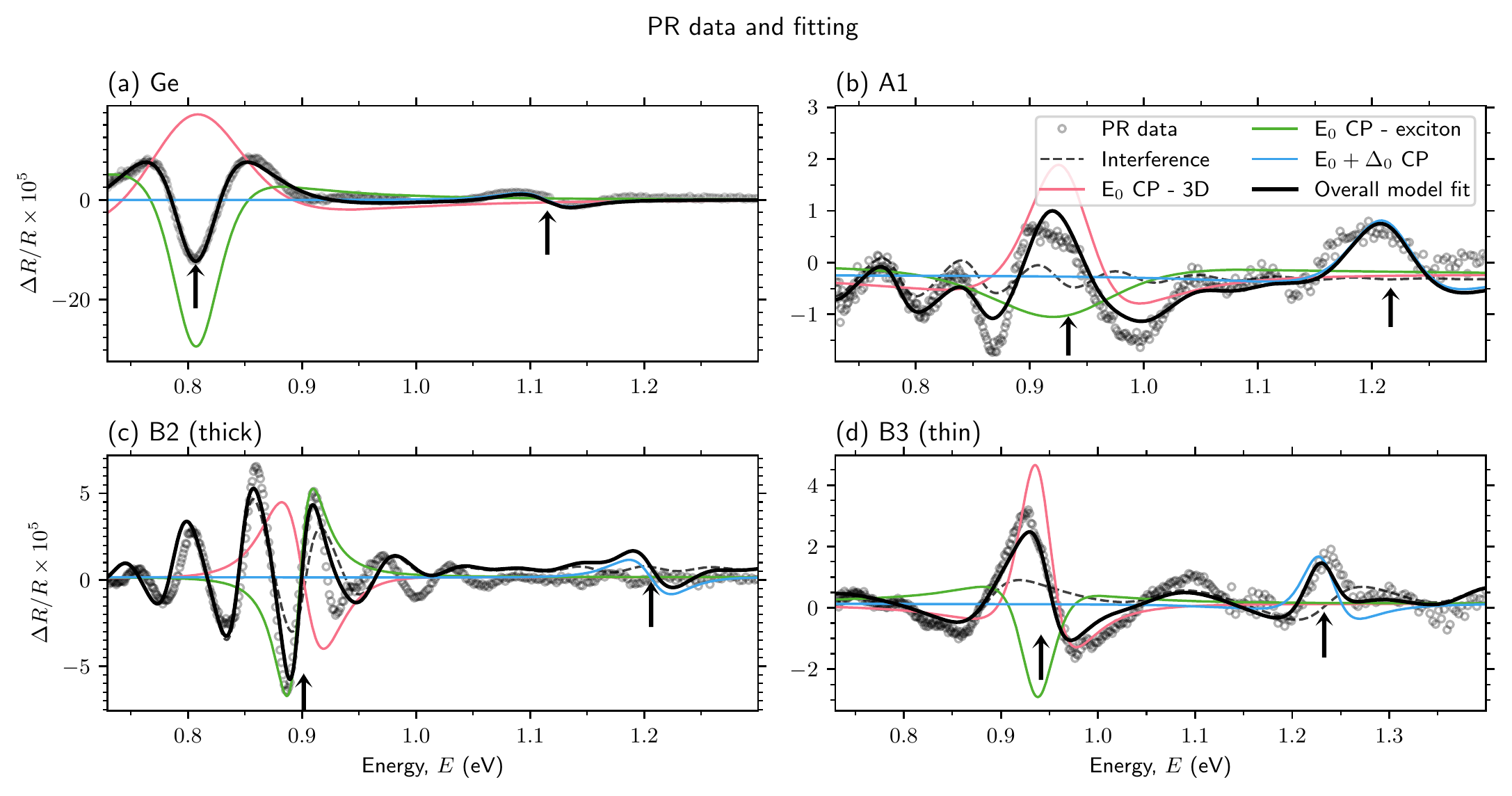}
	\caption{Measured (open black circles) and fitted (solid black lines) low-energy PR spectra for (a) a Ge reference sample and samples (b) A1, (c) B2 (thick), and (d) B3 (thin) (cf.~Table~\ref{tab:samples}).  The left- and right-hand vertical arrows respectively denote the $E_{0}$ and $E_{0} + \Delta_{0}$ transition energies extracted based on the PR fit. The data for Ge was fitted directly to Eq.~\eqref{eq:3DPR}, while fits to the Si$_{x}$Ge$_{1-x-y}$Sn$_{y}$ data were performed using the combined interference/critical point model described in Section~\ref{sec:methodology_photoreflectance}. Dashed red, green and blue lines respectively show the best-fit 3D $M_{0}$-type $E_{0}$ (CB-VB), excitonic, and $E_{0} + \Delta_{0}$ (CB-SO) contributions to the PR fits. Dashed black lines show the thin-film interference contribution to the PR fits for the Si$_{x}$Ge$_{1-x-y}$Sn$_{y}$ data.}
	\label{fig:PR_allfits}
\end{figure*}

Using Eq.~\eqref{eq:weighted_average_energy} therefore allows to compute averaged Si and Sn composition dependent shifts to the alloy $\Gamma$-, L- and X-point CB edge energies, which correspond respectively to the weighted average energies associated with the Ge $\Gamma_{7c}$, L$_{6c}$ and X$_{5c}$ character distributions shown in green, red and blue in Fig.~\ref{fig:cb_character}. Applying the same procedure to compute the Ge $\Gamma_{8v}$ (HH/LH) and $\Gamma_{7v}$ (SO) character allows us to track the evolution of the alloy VB states, and hence to compute the evolution of the direct and indirect alloy band gaps with composition for comparison to experiment in Sec \ref{sec:results_theory_vs_experiment}, below.

% Section 4.3 - Photoreflectance: evolution of direct band gap

\subsection{Photoreflectance: evolution of direct band gap}
\label{sec:results_experiment_pr}

The open black circles in Fig.~\ref{fig:PR_allfits}(a) show the measured PR spectrum for a pure Ge (substrate) reference sample. We note the presence of a critical point feature with a sharp minimum at energy $\approx 0.8$ eV, and an additional feature at $\approx 1.1$ eV, which correspond respectively to the expected direct $\Gamma_{7c}$-$\Gamma_{8v}$ (CB-VB) transition at energy $E_{0}$, and the $\Gamma_{7c}$-$\Gamma_{7c}$ (CB-SO) inter-band transition at energy $E_{0} + \Delta_{0}$ \cite{Adachi1988}. These data were fitted using Eq.~\eqref{eq:3DPR} (solid black line), including critical points for $E_{0}$ and $E_{0} + \Delta_{0}$ transitions, as well as the excitonic contribution at energy $E_{\scalebox{0.7}{\text{ex}}}$ close to $E_{0}$. Previously reported exciton binding energy values for Ge are close to 2 meV \cite{Yin1995}, and so the exciton binding energy was fixed to this value during the fit -- i.e.~$E_{\scalebox{0.7}{\text{ex}}} - E_{0} = -2$ meV. The individual CB-VB, excitonic, and CB-SO contributions to the total PR fit are respectively shown in Fig.~\ref{fig:PR_allfits}(a) using solid red, green and blue lines.

% Figure 6

\begin{figure*}[t!]
	\centering
	\includegraphics[width=0.95\textwidth]{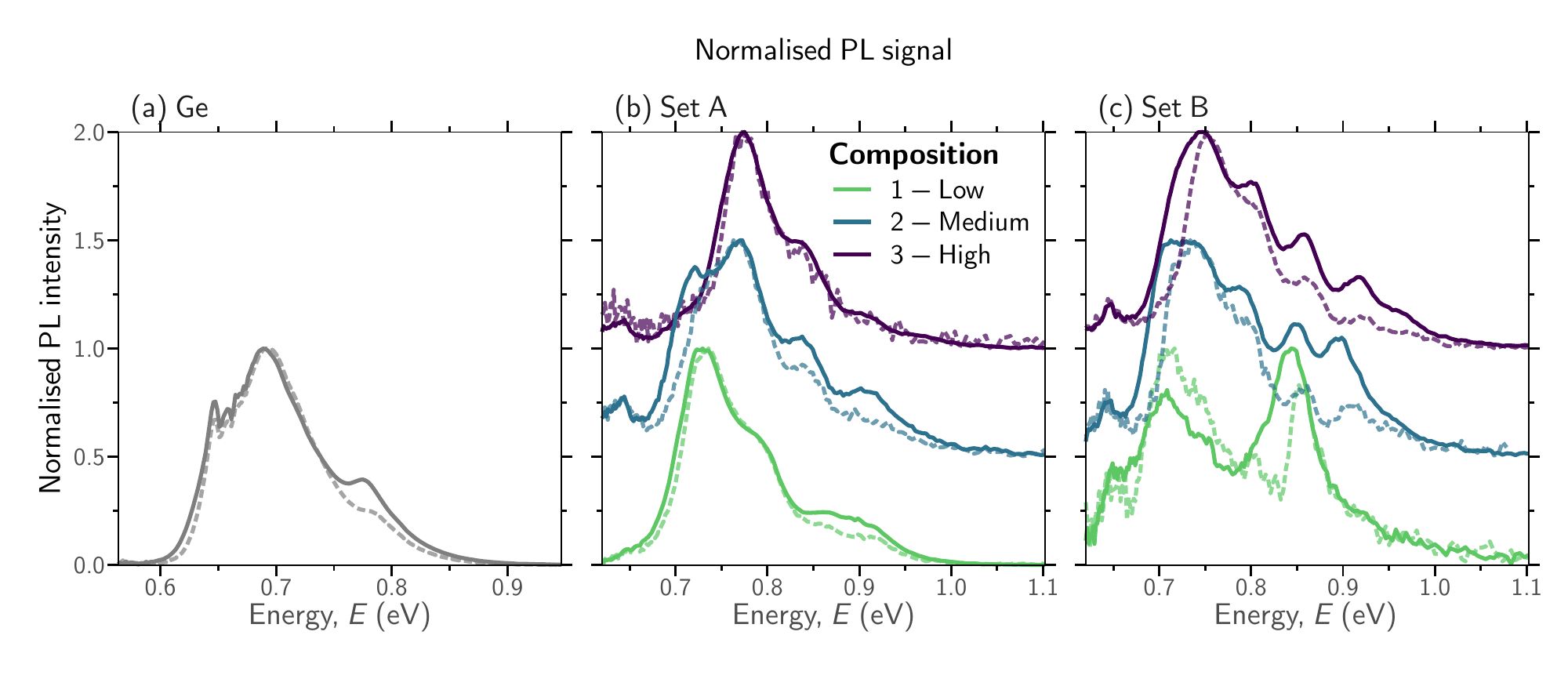}
	\caption{Measured PL spectra of (a) Ge, (b) Si$_{x}$Ge$_{1-x-y}$Sn$_{y}$ alloy sample set A, and (c) Si$_{x}$Ge$_{1-x-y}$Sn$_{y}$ alloy sample set B, at low (dashed lines) and high (solid lines) incident laser power. In (b) and (c) green, blue and purple lines respectively show the measured PL spectra in order of increasing Si and Sn composition (cf.~Table~\ref{tab:samples}). The spectra for the medium and high composition samples are respectively rigidly shifted by +0.5 and +1, to enhance visibility. In (c), measurements are shown only for ``thick'' set B samples. All PL spectra are normalized with respect to their peak luminescence intensity.}
	\label{fig:SiGeSnPLpower}
\end{figure*}

Fitting of the features corresponding to the lowest-energy direct transition in the measured PR spectra for the Si$_{x}$Ge$_{1-x-y}$Sn$_{y}$ samples was performed using the combined interference/critical point model described in Section~\ref{sec:methodology_photoreflectance}, to account for thin-film interference effects in addition to changes in reflectance associated with the critical point feature of interest. Open black circles and solid black lines in Fig.~\ref{fig:PR_allfits} respectively show the measured PR spectra and best model fit for the Si$_{x}$Ge$_{1-x-y}$Sn$_{y}$ samples (b) A1, (c) B2 (thick), and (d) B3 (thin). The combined critical point/interference model was fit using a differential evolution algorithm and, due to the large number of model parameters, running the model multiple times produced slightly different results. However, the key parameters of interest ($E_{0}$ and $\Delta_{0}$) were found to display minimal variation across multiple runs of the fitting algorithm \cite{supmat}. Again, the CB-VB, excitonic, and CB-SO contributions to the overall fit are shown using solid red, green and blue lines for each alloy sample in Figs.~\ref{fig:PR_allfits}(b) --~\ref{fig:PR_allfits}(d), while dashed black lines show the contribution to $\frac{ \Delta R }{ R }$ due to thin-film interference. The best-fit values of $E_{0}$ and $\Delta_{0}$ for all samples are listed in Table~\ref{tab:allfits}. The direct transition energy $E_{0}$ was observed to blueshift strongly with increasing Si and Sn composition, while the VB spin-orbit splitting energy $\Delta_{0}$ remains close to the value 0.308 eV measured for pure Ge. For sample set B, the measured PR data for the ``thick'' and ``thin'' samples were fitted to independently, yet yielded the same average $E_{0}$ energy to within 4 meV. This confirms that the samples have nominally equal Si and Sn compositions at different thicknesses (cf.~Table~\ref{tab:samples}), as expected based on structural characterisation experiments \cite{Pearce2021}. In addition, the best-fit $E_{0}$ values are in good quantitative agreement with values obtained via spectroscopic ellipsometry measurements performed on the same set of samples \cite{Pearce2021}.

% Section 4.4 - Photoluminescence: evolution of indirect band gap

\subsection{Photoluminescence: evolution of indirect band gap}
\label{sec:results_experiment_pl}

\begin{figure*}[t!]
	\centering
	\includegraphics[width=\textwidth]{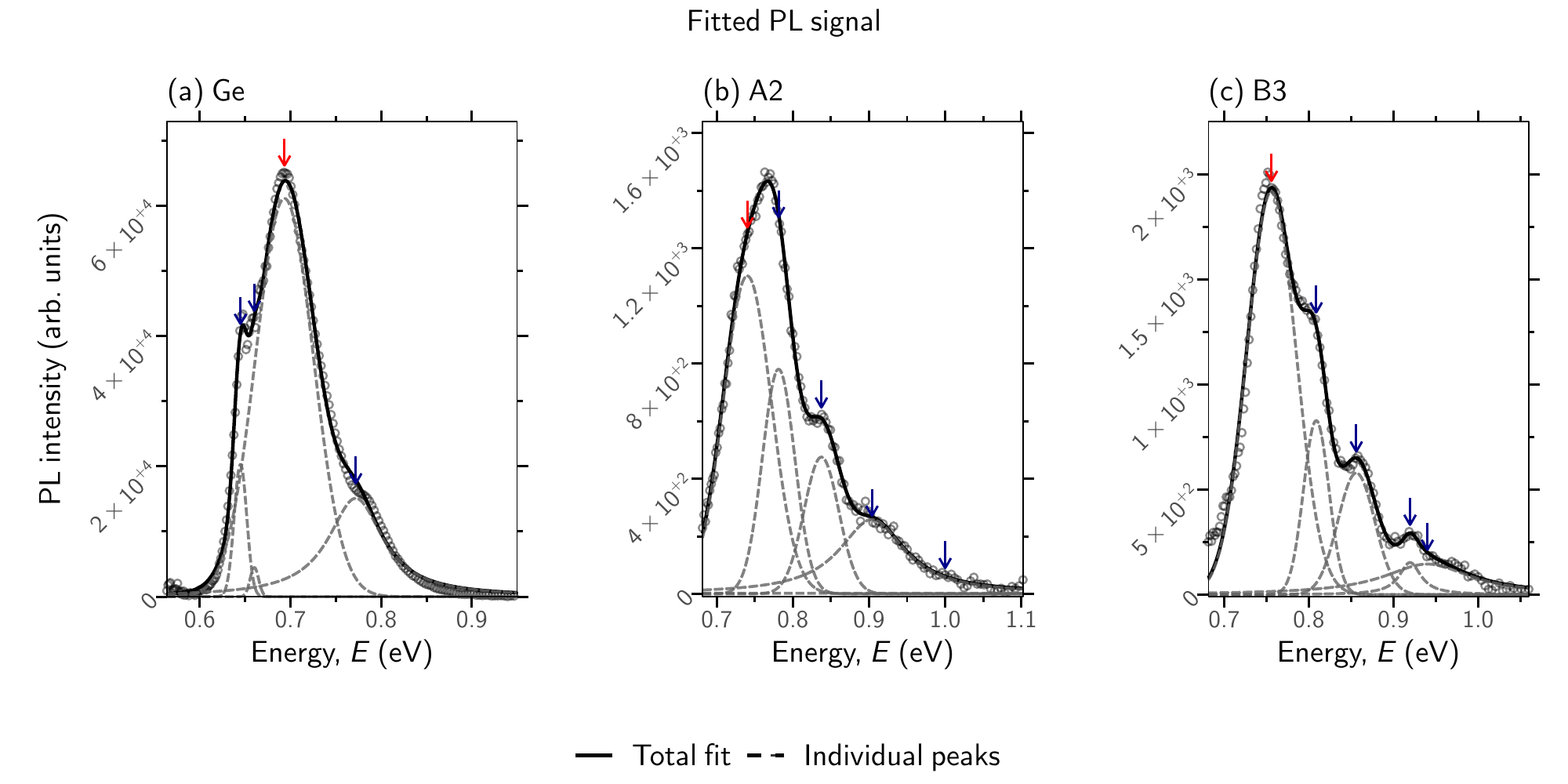}
	\caption{Illustration of the fits to the PL data at low power using Gaussian and Lorentzian line shapes. The total fit to the data and individual peaks are shown for (a) the Ge reference sample, (b) sample A2 and (c) the thicker B3 sample. The arrows indicate the individual peak locations; the red arrow in each case indicates the peak with the highest intensity, which was taken to correspond to LA phonon absorption by analogy with the PL spectrum of Ge.}
	\label{fig:fitshow}
\end{figure*}

PL spectra for the six Si$_{x}$Ge$_{1-x-y}$Sn$_{y}$ alloy samples are shown in Figs.~\ref{fig:SiGeSnPLpower}(b) and (c), where measured PL spectra for sample set B are shown only for the thicker samples due to their observed higher luminescence intensity. The presence of multiple peaks in the measured power-dependent PL spectra for each sample, fitted as shown in Fig. \ref{fig:fitshow}, hinders the straightforward determination of the indirect band gap from the PL data. The origin of these peaks could be phonon-assisted emission involving either L- or X-point CB and $\Gamma$-point VB states (akin respectively to the fundamental band gap of Ge or Si), or direct $\Gamma$-point transitions (akin to the direct $\Gamma_{7c}$-$\Gamma_{8v}$ transition in Ge). Also possible are peaks corresponding to the spectral broadening of the emission spectra predicted by our large-supercell TB calculations of the alloy CB state character, as well as recombination involving localized states associated with native defects. Contributions to the PL spectra of a given sample are also likely to consist of several distinct peaks corresponding to different phonon-assisted transitions, as is the case for Ge (cf.~Fig.~\ref{fig:SiGeSnPLpower}(a) and Ref.~\cite{Lieten2012}). Here, the power dependence of the observed room-temperature PL indicates that it is due to emission from the bulk rather than localized features or defects \cite{supmat}. Our theoretical calculations indicate that, across the composition range under investigation, the Si-like X$_{5c}$-$\Gamma_{8v}$ transition energy of $\gtrsim 1$ eV is significantly higher than the direct $E_{0}$ transition, making it unlikely that this transition contributes significantly to the PL of any of the Si$_{x}$Ge$_{1-x-y}$Sn$_{y}$ samples.

% Figure 7

% Table 2

\begin{table}[]
\caption{\label{tab:allfits}Direct transition energy $E_{0}$ and VB spin-orbit splitting energy $\Delta_{0}$ extracted from fits to measured room temperature PR spectra, and fundamental (indirect) band gap $E_{\protect\scalebox{0.7}{\text{ind}}}$ extracted from fits to measured room temperature PL spectra. Listed values of $E_{0}$ and $\Delta_{0}$ are averages obtained from 10 different runs of the PR fitting algorithm \cite{supmat}. Listed uncertainties are the associated standard deviations. PL measurements were performed only for the ``thick'' ($d \approx 2$ \si{\micro}m) samples from set B (cf.~Table~\ref{tab:samples}).}
    \begin{ruledtabular}
        \begin{tabular}{cccc}
                   & PR                & PR                & PL                                    \\ 
        Sample     & $E_{0}$ (eV)      & $\Delta_{0}$ (eV) & $E_{\scalebox{0.7}{\text{ind}}}$ (eV) \\ \hline
        Ge         & 0.807             & 0.308             & 0.666                                 \\
        \hline
        A1         & $0.941 \pm 0.007$ & $0.28 \pm 0.01$   & 0.705                                 \\
        A2         & $1.000 \pm 0.020$ & $0.27 \pm 0.02$   & 0.713                                 \\
        A3         & $1.010 \pm 0.002$ & $0.30 \pm 0.01$   & 0.745                                 \\
        \hline
        \multicolumn{1}{l}{B1 (thin)}  & $0.870 \pm 0.006$ & $0.29 \pm 0.01$   & -----                                 \\
        \multicolumn{1}{l}{B1 (thick)} & $0.870 \pm 0.003$ & $0.29 \pm 0.01$   & 0.690                                 \\
        \hline
        \multicolumn{1}{l}{B2 (thin)}  & $0.910 \pm 0.003$ & $0.30 \pm 0.01$   & -----                                 \\
        \multicolumn{1}{l}{B2 (thick)} & $0.910 \pm 0.020$ & $0.29 \pm 0.01$   & 0.711                                 \\
        \hline
        \multicolumn{1}{l}{B3 (thin)}  & $0.950 \pm 0.010$ & $0.29 \pm 0.01$   & -----                                 \\
        \multicolumn{1}{l}{B3 (thick)} & $0.950 \pm 0.008$ & $0.30 \pm 0.01$   & 0.726                                 \\
        \end{tabular}
    \end{ruledtabular}
\end{table}

Thus, the most likely origin of the multiple PL peaks in the measured Si$_{x}$Ge$_{1-x-y}$Sn$_{y}$ room temperature PL spectra is different phonon-assisted L$_{6c}$-$\Gamma_{8v}$ recombination processes, in combination with emission from the direct $\Gamma_{7c}$-$\Gamma_{8v}$ gap, with the latter appearing as a high-energy tail. For the indirect gap emission processes we do not expect that any individual PL peak energy corresponds directly to the fundamental band gap, as the energies of these features will be offset by the energy of the emitted or absorbed phonon, with the observed phonon-assisted PL peak energy being lower (higher) in energy than the associated indirect band gap for a transition assisted by phonon emission (absorption). There is therefore some difficulty in associating the observed low-energy PL peaks directly to specific phonon modes, firstly since the loss of translational symmetry in the alloy removes the requirement to conserve crystal momentum, allowing a large number of phonons modes to contribute to indirect recombination, and secondly due to a lack of information on how phonon energies vary in disordered Si$_{x}$Ge$_{1-x-y}$Sn$_{y}$ alloys compared to Ge. Overall, we note that there does not appear to be a direct correspondence between the measured energy offsets between the phonon-assisted PL peak energies and associated indirect fundamental band gap for the Si$_{x}$Ge$_{1-x-y}$Sn$_{y}$ alloy samples compared to the Ge reference sample. This makes it challenging to associate specific phonon modes to observed indirect PL peaks. Therefore, the room-temperature band gap of the Si$_{x}$Ge$_{1-x-y}$Sn$_{y}$ samples was estimated by identifying the highest-intensity peak in the Ge room-temperature PL spectrum, which corresponds to an indirect L$_{6c}$-$\Gamma_{8v}$ transition assisted by absorption of a longitudinal acoustic phonon having energy 27 meV \cite{Madelung2002}. In order to consistently estimate the fundamental indirect band gap energy, it was assumed that the highest-intensity peak in the measured PL spectrum for each Si$_{x}$Ge$_{1-x-y}$Sn$_{y}$ sample was also associated with absorption of a phonon of energy 27 meV. The resulting estimated indirect band gaps are listed in Table \ref{tab:allfits}, and plotted in \ref{fig:theory_vs_experiment}(b) using closed red squares, demonstrated a linear blueshift with increasing Si and Sn composition.

% Section 4.5 - Theory vs.~experiment: direct and indirect band gaps

\subsection{Theory vs.~experiment: direct and indirect band gaps}
\label{sec:results_theory_vs_experiment}

\begin{figure}[]
	\centering
	\includegraphics[width=0.9\columnwidth]{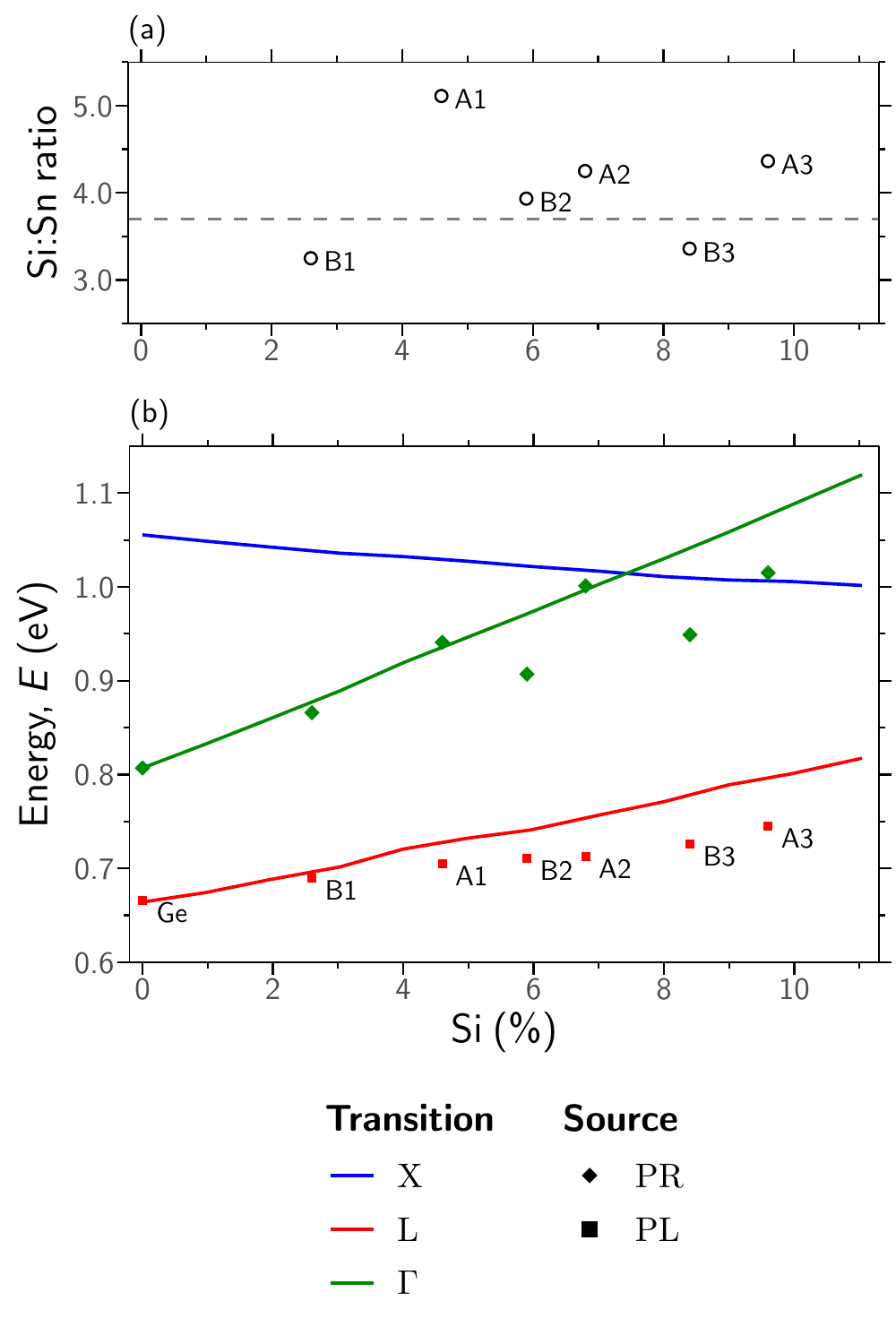}
	\caption{(a) SEM-EDX measured Si to Sn composition ratios $x$:$y$ for the samples investigated (cf.~Table~\ref{tab:samples}). (b) Comparison of the transition energies extracted from experimental measurements (points) and calculated using TB method (lines), showing data for the indirect band gap from PL measurements (red squares), and for the direct transition from PR measurements (green diamonds). The theoretical results are the calculated weighted average direct $\Gamma_{7c}$-$\Gamma_{8v}$ (green), and indirect L$_{6c}$-$\Gamma_{8v}$ (red) and X$_{5c}$-$\Gamma_{8v}$ (blue) transition energies.}
	\label{fig:theory_vs_experiment}
\end{figure}

The results of our PR (closed green diamonds) and PL (closed red squares) measurements of the direct and indirect band gaps are summarized in Fig.~\ref{fig:theory_vs_experiment}(b). Following the procedure described in Section~\ref{sec:results_theory_tb}, we have used the TB method to compute the composition dependence of the weighted average direct and indirect band gaps of lattice-matched Si$_{x}$Ge$_{1-x-y}$Sn$_{y}$. The results of these calculations are shown in Fig.~\ref{fig:theory_vs_experiment}(b) using open triangles for the direct $\Gamma_{7c}$-$\Gamma_{8v}$ (green), and indirect L$_{6c}$-$\Gamma_{8v}$ (red) and X$_{5c}$-$\Gamma_{8v}$ (blue) band gaps. In order to compare our zero temperature theoretical calculations and room temperature experimental measurements, we have rigidly redshifted the calculated band gaps so that the calculated direct $\Gamma_{7c}$-$\Gamma_{8v}$ band gap of Ge aligns with the value of 0.807 eV extracted from our PR measurements (cf.~Table~\ref{tab:allfits}). Applying this same redshift rigidly to all calculated band gaps then allows the predicted composition dependence to be quantitatively compared to experiment.

% Figure 8

In comparing our theoretical calculations and experimental measurements, we note three important considerations. Firstly, our theoretical calculations are for nominally lattice-matched alloy supercells, where the supercell lattice constant conforms very closely to that Ge, while our CVD-grown samples possess composition ratios which deviate from the ideal lattice-matched Si to Sn composition ratio, as shown in Fig.~\ref{fig:theory_vs_experiment}(a). As a result, a given Si composition in Fig.~\ref{fig:theory_vs_experiment}(b) does not imply equal Sn compositions in theory and experiment. The measured lattice constants for these samples suggest lattice mismatch of $\leq 0.03$\% with respect to Ge \cite{Pearce2021}, and thus strain is not expected to impact the measured transition energies. Secondly, we recall that the CB and VB energies used to calculate the theoretical transition energies are computed as a weighted average of the distribution of the corresponding Ge $\Gamma$-, L- or X-point Bloch character over individual alloy supercell states. These weighted average energy gaps therefore display virtual crystal-like behavior -- i.e. approximately linear variation with composition, with bowing of individual state energies averaged out -- which can only be expected to describe the composition dependence of a given transition energy on average in terms of its contributions from a multiplicity of states possessing an admixture of $\Gamma$-, L- or X-character and spread over a range of energies in the alloy band structure. As described in Section~\ref{sec:methodology_theoretical}, the allowed band mixing in a given alloy supercell is determined in part by the supercell size, which controls the number of states that fold to a given supercell wave vector $\textbf{K}$. We therefore expect that our calculated weighted average band gaps will tend to overestimate the corresponding transition energies measured in experiment, since only a limited number of states fold back to $\textbf{K} = 0$ close in energy to the CB edge in a given alloy supercell. Thirdly, as discussed in Section \ref{sec:results_experiment_pl}, the presence of multiple peaks in the PL spectra, associated with different phonon-assisted emission pathways, means there is uncertainty associated assigning a single energy to the measured indirect fundamental band gap.

Bearing these potential sources of error in mind, we nonetheless note good overall agreement between theory and experiment in Fig.~\ref{fig:theory_vs_experiment}(b). Our calculated increase of 28.4 meV per \% Si of the direct $\Gamma_{7c}$-$\Gamma_{8v}$ transition (open red triangles) corresponds well to the $E_{0}$ values extracted from the PR measurements (closed red diamonds) for samples A1, A2 and B1, but overestimates the direct transition energy measured for samples A3, B2 and B3. Similarly, our calculated increase of 14.1 meV per \% Si of the indirect L$_{6c}$-$\Gamma_{8v}$ band gap slightly overestimates, but corresponds well to, the fundamental band gap identified from our PL measurements. As predicted by theory, the VB spin-orbit splitting energy $\Delta_{0}$ does vary significantly across the composition range. Based on this analysis we conclude that lattice-matched Si$_{x}$Ge$_{1-x-y}$Sn$_{y}$ retains an indirect fundamental band gap, associated primarily with an L-point CB minimum, albeit with theoretical calculations suggesting that band hybridisation effects result in this fundamental band gap having a hybridized nature, possessing an admixture of Ge $\Gamma_{7c}$ character that should allow for (weak, spectrally broad) optical absorption at energies below the weighted average direct band gap indicated in Fig.~\ref{fig:theory_vs_experiment}. Finally, our calculated reduction of 4.6 meV per \% Si for the X$_{5c}$-$\Gamma_{8v}$ band gap results in a crossover from L-$\Gamma$-X to L-X-$\Gamma$ CB edge energy ordering for $x \gtrsim 7.5$\%. We note that this is predicted to occur in the composition range in which the direct $E_{0}$ transition is close to 1 eV in magnitude at room temperature, in accordance with the band ordering identified in our DFT effective band structure calculations (cf.~Fig.~\ref{fig:unfolded_bands}).

For multi-junction solar cell applications, the indirect nature of the fundamental band gap of Si$_{x}$Ge$_{1-x-y}$Sn$_{y}$ alloys having a direct $\Gamma_{7c}$-$\Gamma_{8v}$ transition energy close to 1 eV ($x \approx 8.0$\%, $y \approx 2.2$\%) is expected to have a detrimental impact on the cell voltage. For application as a 1 eV junction, where lattice-matched Si$_{x}$Ge$_{1-x-y}$Sn$_{y}$ is grown epitaxially with thicknesses up to several $\mu$m, the direct $\Gamma_{7c}$-$\Gamma_{8v}$ transition will dominate the optical absorption. This offers advantages for current-matching in three- or four-junction solar cells compared to a cell having a pure Ge junction. However, the presence of the lower-energy indirect L$_{6c}$-$\Gamma_{8v}$ band gap will allow carriers to thermalize to lower energy CB states after being absorbed across the $\Gamma_{7c}$-$\Gamma_{8v}$ transition, thus reducing the voltage that can be extracted from the cell compared to a material having a 1 eV direct fundamental band gap. Nonetheless, since the fundamental indirect band gap is increased in lattice-matched Si$_{x}$Ge$_{1-x-y}$Sn$_{y}$ compared that of Ge, Si$_{x}$Ge$_{1-x-y}$Sn$_{y}$ sub-cells should be able to achieve a higher voltage, and thus an overall improved triple-junction cell efficiency, compared to an equivalent cell employing a pure Ge junction.

%%%%%%%%%%%%%%%%%%%%%%%%%%%%%%%%
%%%% Section 5: Conclusions %%%%
%%%%%%%%%%%%%%%%%%%%%%%%%%%%%%%%

\section{Conclusions}
\label{sec:conclusions}

% High-level summary

In summary, we have presented a combined experimental and theoretical analysis of ternary Si$_{x}$Ge$_{1-x-y}$Sn$_{y}$ alloys lattice-matched to Ge and GaAs. Our experimental measurements quantified the evolution of the optical properties with alloy composition, focusing on trends in the direct and indirect band gaps, while our theoretical calculations elucidated these trends in terms of the evolution of the alloy electronic structure. Our results provide new insight into the Si$_{x}$Ge$_{1-x-y}$Sn$_{y}$ alloy band structure, demonstrating that its evolution is characterized primarily by a perturbed CB structure that undergoes rapid change with increasing Si and Sn composition.

Experimentally, we reported room temperature PR and PL spectroscopic measurements, which were respectively employed to track the evolution of the direct and indirect band gaps as a function of alloy composition. Theoretically, we reported first principles and semi-empirical calculations of the alloy electronic structure. Effective (unfolded) first principles band structure calculations for lattice-matched Si$_{x}$Ge$_{1-x-y}$Sn$_{y}$ highlighted the presence of Si- and Sn-induced alloy band mixing, similar to that observed in recent calculations for binary Ge$_{1-y}$Sn$_{y}$ alloys. We predicted theoretically and confirmed via photoluminescence measurements that lattice-matched Si$_{x}$Ge$_{1-x-y}$Sn$_{y}$ alloys retain an indirect fundamental band gap, and further identified that the electronic structure evolution at low Si and Sn composition is strongly influenced by the impact of alloy band mixing on the CB. This alloy band mixing, combined with a breakdown in \textbf{k}-selection due to a loss of translational symmetry in a disordered alloy, generates significant energetic broadening of CB edge states. Large-scale semi-empirical calculations further elucidated this behavior, demonstrating that alloying results in a distribution of direct-gap (Ge $\Gamma_{7c}$) character across a multiplicity of alloy CB states. Taken together, our theoretical calculations and experimental measurements suggest that the observed strong inhomogeneous broadening of optical spectra is then an intrinsic property of substitutional Si$_{x}$Ge$_{1-x-y}$Sn$_{y}$ alloys. Our experimental measurements and theoretical calculations confirm the previously demonstrated blueshift of the direct band gap with increasing Si composition, suggesting that it increases at a rate of $\approx 28$ meV per \% Si in a lattice-matched alloy.

Our results confirm that Si$_{x}$Ge$_{1-x-y}$Sn$_{y}$ alloys lattice-matched to Ge possess an indirect fundamental band gap which, for $E_{0}$ $\approx 1$ eV, lies 250-300 meV below the $\Gamma_{7c}$-$\Gamma_{8v}$ transition in energy, which theoretical calculations identify as resulting from alloy CB edge states being primarily derived from the L$_{6c}$ CB minimum of the Ge host matrix semiconductor. While the presence of an indirect fundamental band gap is expected to adversely affect the voltage of a Si$_{x}$Ge$_{1-x-y}$Sn$_{y}$ solar cell compared to a material having a 1 eV direct band gap, the improved potential for current-matching in lattice-matched architectures and an expected increase in voltage compared to an equivalent Ge cell make Si$_{x}$Ge$_{1-x-y}$Sn$_{y}$ alloys a promising candidate material for multi-junction solar cells.

%%%%%%%%%%%%%%%%%%%%%%%%%%
%%%% Acknowledgements %%%%
%%%%%%%%%%%%%%%%%%%%%%%%%%

\section*{Acknowledgements}

This work was supported by the Engineering and Physical Sciences Research Council, U.K. (EPSRC; via a CASE Studentship, held by P.P.~and sponsored by IQE plc.), by the National University of Ireland (NUI; via the Post-Doctoral Fellowship in the Sciences, held by C.A.B.), by Science Foundation Ireland (SFI; project no.~15/IA/3082), and by the Royal Society (via an Industry Fellowship, held by N.J.E-D.). This work was carried out in part during research visits to Imperial College London, U.K. by C.A.B., supported by a Charlemont Grant from the Royal Irish Academy. The authors thank S. W. Ong (National University of Singapore) for performing the SEM-EDX composition measurements.

%%%%%%%%%%%%%%%%%%%%%%%%%%%%%%%
%%%% Data access statement %%%%
%%%%%%%%%%%%%%%%%%%%%%%%%%%%%%%

% \section*{Data access}

% The data associated with this work are openly available, and can be accessed via Ref.~\onlinecite{Pearce_SiGeSn_2020_data}.

%%%%%%%%%%%%%%%%%%%%
%%%% References %%%%
%%%%%%%%%%%%%%%%%%%%

% References which appear in the SI need to appear here, but don't want to cite them in the main text

\nocite{Kallergi1990}
\nocite{Huang1989}
\nocite{Adachi2005}
\nocite{Klingenstein1978}
\nocite{VanDriel1976}
\nocite{Schmidt1992}
\nocite{Rathgeber_ACMTMS_2016}

\bibliographystyle{apsrev} % Using the Physical Review style for references
\bibliography     {main.bib} % The bibliography is contained in the file: SiGeSn.bib

\end{document}

\widetext
\title{Supplemental Material for ``Electronic and optical properties of\\Si$_{x}$Ge$_{1-x-y}$Sn$_{y}$ alloys lattice-matched to Ge''}

%%%%%%%%%%%%%%%%%%%%%%%%%%%%%%%%%%%%%%%%
%%%% Authors, affiliations and date %%%%
%%%%%%%%%%%%%%%%%%%%%%%%%%%%%%%%%%%%%%%%

\author{Phoebe M. Pearce}
\email{p.pearce@unsw.edu.au} % Email address must be kept before affiliation(s)
\affiliation{Department of Physics, Imperial College London, South Kensington, London SW7 2AZ, United Kingdom}

\author{Christopher A.~Broderick}
\email{c.broderick@umail.ucc.ie} % Email address must be kept before affiliation(s)
\affiliation{Tyndall National Institute, University College Cork, Lee Maltings, Dyke Parade, Cork T12 R5CP, Ireland}
\affiliation{Department of Physics, University College Cork, Cork T12 YN60, Ireland}

\author{Michael P. Nielsen}
\affiliation{School of Photovoltaic and Renewable Engineering, University of New South Wales, Sydney, New South Wales 2052, Australia}

\author{Andrew D.~Johnson}
\affiliation{IQE plc., Pascal Close, St. Mellons, Cardiff CF3 0LW, United Kingdom}

\author{Nicholas J.~Ekins-Daukes}
\affiliation{Department of Physics, Imperial College London, South Kensington, London SW7 2AZ, United Kingdom}
\affiliation{School of Photovoltaic and Renewable Engineering, University of New South Wales, Sydney, New South Wales 2052, Australia}

\date{\today}

% \keywords{0.0X}
% \pacs{0.0X}

\maketitle

%%%%%%%%%%%%%%%%%%%%%%%%%%%%%%%%%%%%%%%%%%%%%%
%%%% Section 1: Experimental measurements %%%%
%%%%%%%%%%%%%%%%%%%%%%%%%%%%%%%%%%%%%%%%%%%%%%

\section{Photoreflectance data fitting}

\label{sec:experiment_pr}

\begin{figure}
	\centering
	\includegraphics[width=\textwidth]{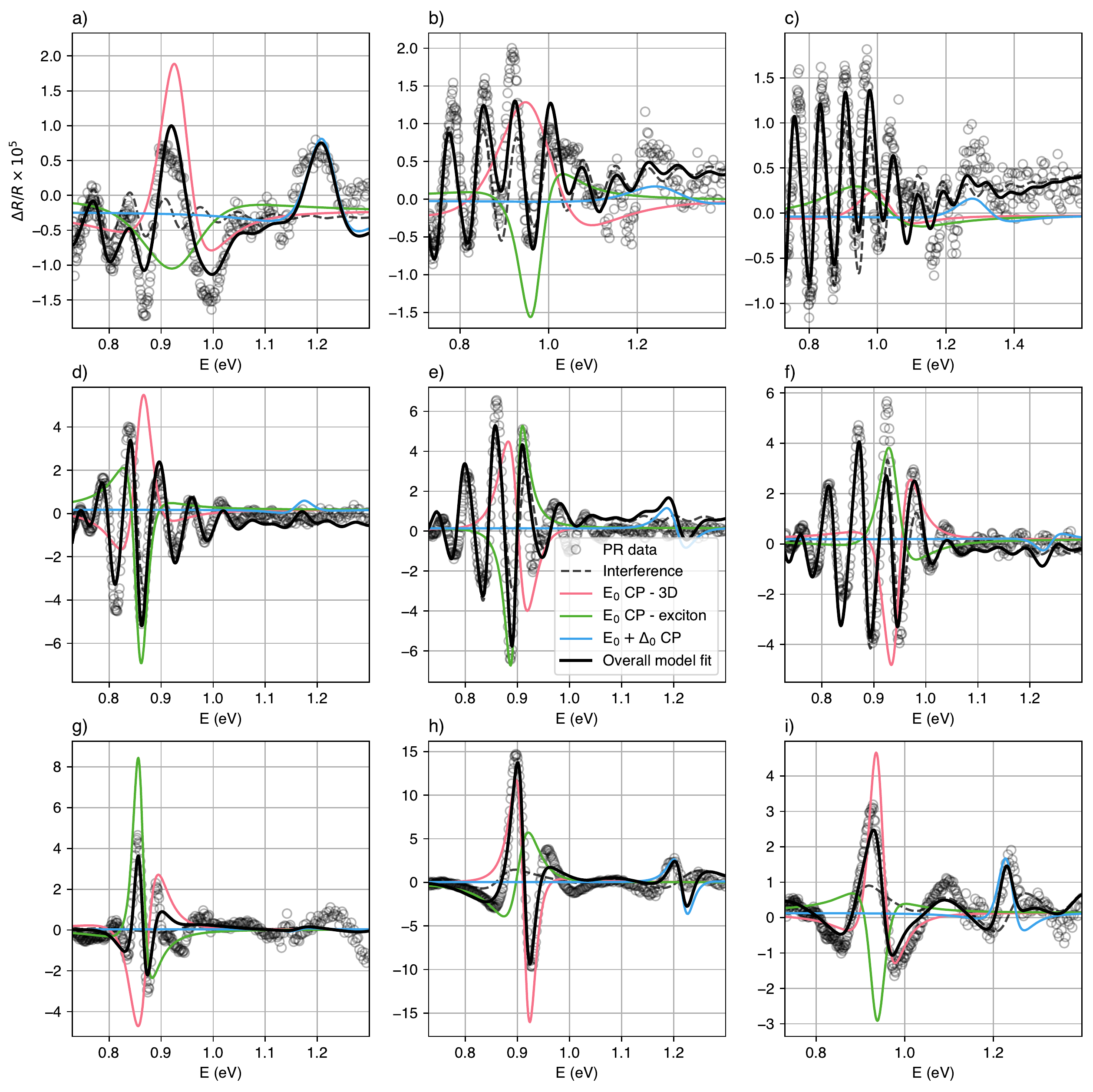}
	\caption{Best fits (out of 10 runs of the differential evolution model fitting) to the PR data of the  samples from a)-c) Set A, from lowest to highest compositio; d)-f) $\sim$ 2 \si{\micro}m Set B samples from lowest to highest composition; and g)-i) $\sim$ 400 nm Set B samples from lowest to highest composition, using the combined interference/critical point model. The overall model fit (black line) is shown in addition to contributions from thin-film interference, the $E_0$ critical point (plus excitonic contribution) and the $E_0 + \Delta_0$ critical point.}
	\label{fig:PR_allfits}
\end{figure}

Assuming the anomalous features observed in the PR of the  samples (see Fig. \ref{fig:PR_allfits}) are interference fringes due to reﬂection at the buried InGaAs interface, the period of oscillations of the fringes is determined by the Si$_{x}$Ge$_{1-x-y}$Sn$_{y}$ layer thickness, while the envelope shape of the oscillations depends on the modulation of the refractive index by the pump beam. While Kallergi \cite{Kallergi1990} and Huang \cite{Huang1989} attribute the interference to modulation of the refractive index of doped GaAs substrates, rather than modulation of the refractive index of the epi-layer, this explanation is unlikely for the Si$_{x}$Ge$_{1-x-y}$Sn$_{y}$ samples as the thickness of the samples ($\approx 1.8-2$ \si{\micro}m) means that very few ($<$1\%) of the high energy photons from the pump beam (532 nm laser) incident on the sample are able to reach the GaAs substrate (Set A) or InGaAs buffer layer (Set B). However, the model presented in [\onlinecite{Kallergi1990}] and [\onlinecite{Huang1989}] can be adapted relatively simply to assume modulation of the epilayer refractive index rather than the substrate. 

The photoreflectance signal $\Delta R/R$ is the relative change in the reflectivity with and without the pump beam, so  $\Delta R/R = (R'-R)/R$ where $R'$ is the modulated reflectivity and $R$ is the unmodulated reflectivity. For instance, for Set A, considering the three layers in the epitaxial structure (Si$_{x}$Ge$_{1-x-y}$Sn$_{y}$, Ge seed layer, and GaAs substrate), the reflectivity is a function of the relevant thicknesses and refractive indices: the unmodulated reflectivity $R(E,n_{SiGeSn},n_{Ge},n_{(In)GaAs},d_{SiGeSn}, d_{Ge})$ and the modulated reflectivity $R'(E, n_{SiGeSn} + \delta n_{SiGeSn}, n_{Ge} + \delta n_{Ge},n_{GaAs} + \delta n_{GaAs}, d_{SiGeSn}, d_{Ge})$ where the $n$ are the complex refractive indices. Assuming the refractive indices of the Ge and GaAs are not significantly affected since very little pump light is able to reach these layers, the functional dependence of the modulated reflectivity simplifies to $R'(E, n_{SiGeSn} + \delta n_{SiGeSn}, n_{Ge},n_{GaAs}, d_{SiGeSn}, d_{Ge})$. Because of the presence of the Ge seed layer, the multi-layer reflectivity model is more complex than in [\onlinecite{Kallergi1990}] and [\onlinecite{Huang1989}], and rather than constructing an analytical expression for the multi-layer interference, Solcore’s TMM optical model \cite{Alonso-Alvarez2018} was used to calculate both $R'$ and $R$, and thus $\frac{\Delta R}{R}$ with and without modulation of the dielectric function of the . The change in the refractive index was expressed as in [\onlinecite{Kallergi1990}]:

\begin{equation}
\delta n_{SiGeSn} \propto (n_{SiGeSn}^2-1)e^{-\gamma E}
\end{equation}

The  thickness, constant of proportionality $A$, and $\gamma$ were fitted to the data ($d_{Ge}$ was kept fixed at 60 nm). 

Optical constants for the different  compositions were obtained from ellipsometry\cite{Pearce_SiGeSn_SE_2021}, and these could be used in this model for $n_{SiGeSn}$. However, this raises an issue, since the location of the critical points which occur in the PR data is what leads to certain values of $n$ and $\kappa$; using externally imposed values of $n$ and $\kappa$ is not consistent with also fitting the location of the critical points. However, it is possible to instead fit the interference fringes and signal from the critical point simultaneously by expressing the optical constants in terms of an analytical model such as the Adachi model\cite{Adachi2005} and fitting the optical constant modulation using the optical constants calculated directly from these analytical expressions, and the photoreflectance signal to their third derivative.

\begin{table}[h]
\caption{\label{tab:PRparams}All parameters used in fitting the photoreflectance data to a model incorporating both the critical point feature and the effects of thin-film interference in the  layer. The ticks in the final two columns indicate whether the parameter affects the fit of the interference fringes, the critical point feature, or both.}
\begin{tabular}{llllllll}
\hline
\textbf{Parameter} & \textbf{Description} & \textbf{Interference fringes} & \textbf{PR signal due to CP} \\ \hline
$E_0$  &		 $E_0$  CP ($\Gamma \rightarrow \Gamma$) centre energy & \checkmark & \checkmark \\
$\Delta_0$	&		 Split-off hole energy & \checkmark & \checkmark \\
$A_0$  &		 Strength of the $E_0$ transition & \checkmark & \\
$\Gamma_{0/\Delta_0}$ & 		Broadening of the $E_0$ transition & \checkmark & \checkmark \\
$\gamma$ & 		Exponent for envelope & \checkmark & \\
$A_n$ & 		Strength of $n$ modulation & \checkmark & \\
$A_{CP, 0}$ &		 Strength of $E_0$ CP signal &  & \checkmark \\
$\Phi_{0/\Delta_0}$ & 		Phase of $E_0$ and $E_0 + \Delta_0$ CP signal & & \checkmark \\
$A_{CP, \Delta_0}$ & 		Strength of $E_0 + \Delta_0$ CP signal & & \checkmark \\
$E_{ex}$ & 	$E_0$ exciton binding energy transition energy & & \checkmark\\
$A_{ex}$ & 		Strength of the exciton PR signal & & \checkmark\\
$\Gamma_{ex}$ & 		Broadening of the exciton PR signal & & \checkmark\\
$\Phi_{ex}$ & 		Phase of the exciton PR signal & & \checkmark\\
$d_{SiGeSn}$ & 	Thickness of the  layer & \checkmark &\\ 
$d_{GaAs}$ & 		Thickness of the GaAs layer (Set B only) & \checkmark &\\ 
$\Phi_{PR}$ & 		Overall phase of the PR measurement & \checkmark & \checkmark \\ 
$C$ & 			Constant offset of measurement & \checkmark & \checkmark\\ 
\end{tabular}

\end{table}

The real and imaginary parts of the dielectric constant, and thus the refractive index $n$, are calculated according to: \cite{Adachi2005}
\begin{equation}
\label{eq:adachin}
\begin{array}{l}
\varepsilon=A E_{0}^{-1.5}\left\{f\left(\chi_{0}\right)+\frac{1}{2}\left(\frac{E_{0}}{E_{0}+\Delta_{0}}\right)^{1.5} f\left(\chi_{\mathrm{so}}\right)\right\} \\

f\left(\chi_{0}\right)=\chi_{0}^{-2}\left[2-\left(1+\chi_{0}\right)^{0.5}-\left(1-\chi_{0}\right)^{0.5}\right] \\
f\left(\chi_{\mathrm{so}}\right)=\chi_{\mathrm{so}}^{-2}\left[2-\left(1+\chi_{\mathrm{so}}\right)^{0.5}-\left(1-\chi_{\mathrm{so}}\right)^{0.5}\right] \\

\chi_{0}=\frac{\hbar \omega+i \Gamma_{0/\Delta_0}}{E_{0}} \\
\chi_{\mathrm{so}}=\frac{\hbar \omega+i \Gamma_{0/\Delta_0}}{E_{0}+\Delta_{0}} \\

n=\Re\{\sqrt{\varepsilon}\} \\

\end{array}
\end{equation}

The change in the refractive index of  due to modulation by the pump beam is given by (using the $n$ from eq. \ref{eq:adachin}):
\begin{equation}
\begin{array}{l}
\delta_{n}=A_{n} e^{\gamma E}\left(n^{2}-1\right) \\
\tilde{n} = n + \delta n \\

\end{array}
\end{equation}

This modulated refractive index is fed into the TMM model of the layer stack to calculate $R'$; $R$ is similarly calculated with the unmodulated refractive index $n$, so that the change in reflectivity $\Delta R = R'-R$ due to interference can be calculated. The interference signal $IS$ is given by $(R'-R)/R$.

The critical point feature in the photoreflectance data is fitted using:

\begin{equation}
\begin{array}{l}
CP = \Re \left[ \frac{A_{CP,0}e^{i\Phi_0}}{(E-E_0+i\Gamma_{0/\Delta_0})^{5/2}} +
\frac{A_{ex}e^{i\Phi_{ex}}}{(E-E_{ex}+i\Gamma_{ex})^{2}} +
\frac{A_{CP, \Delta_0}e^{i\Phi_{\Delta_0}}}{(E-(E_0 + \Delta_0)+i\Gamma_{\Delta_0})^{5/2}} \right]  \\

\end{array}
\end{equation}

Finally, the overall modelled value for the PR signal is the sum of the contributions, plus a constant offset. In addition, a phase factor is fitted (this is due to the phase chosen for the lock-in amplifier during the measurement):

\begin{equation}
    \frac{\Delta R}{R} = \Re\{(CP + IS + C)e^{i \Phi_{PR}}\}
\end{equation}

The symbols used for each parameter, and their physical meaning, are given in Table \ref{tab:PRparams}. 

This model, which fits the full photoreflectance signal, has a large number of parameters to account for the layer thicknesses, peak strength and broadening, phase factors, and degree of optical constant modulation. The differential evolution (DE) optimization technique was used to fit the model to the data for all nine  samples. Since the model has a large number of parameters, the differential evolution was run multiple times for each sample to identify whether consistent results could be obtained between runs of the optimization algorithm. The fits with the lowest root mean square error (RMSE) are shown in Fig. \ref{fig:PR_allfits}. A list of all fitted parameters fitted and the bounds imposed is given in Table \ref{tab:full_PR_fitresults}. This shows that while the standard deviation across 10 iterations of the model fitting procedure was large for some parameters, the error associated with the key parameters $E_0$ and $\Delta_0$ was relatively small.

As the Set B samples were grown after the Set A samples, when the effect of interference of the PR signal had been identified, two versions of each composition were grown; $\sim 2$ \si{\micro}m samples, similar to the Set A samples, and thinner $\sim$ 400 nm  layers were grown as part of Set B. The expectation was that for the thinner samples, the interference fringes would be significantly wider, and the photoreflectance signal arising from the critical points would be more easily identifiable; this can be seen clearly in the data in Fig. \ref{fig:PR_allfits}(g)-(i). The two different thicknesses of Set B samples were fitted independently, but the values for $E_0$ and $\Delta_0$ are in good agreement for each composition, validating the model presented here for simultaneous fitting of the critical points and interference fringes, and the consistency of the composition grown across two different sample thicknesses.

\begin{turnpage}

\begin{table}[]
\setlength{\tabcolsep}{3pt}
\caption{Full results of fits to the photoreflectance data using the combined thin-film interference and critical point model. The meaning of the parameters is given in Table \ref{tab:PRparams}. The fit is run 10 times for each sample; the values given are the mean result across the 10 fitting runs, with the standard deviation given in brackets.}
\label{tab:full_PR_fitresults}
\footnotesize
\begin{tabular}{l|l|l|l|l|l|l|l|l|l|l|l|l|l|l|l}
Sample     & $\mathbf{A_0}$         & $\mathbf{E_0}$               & $\mathbf{\Gamma_0}$           & $\mathbf{\Delta_0}$           & $\mathbf{A_{CP,0}}$         & $\mathbf{\Phi_{0/\Delta_0}}$ & 
$\mathbf{A_{CP, \Delta_0}}$    & $\mathbf{A_{ex}}$        & $\mathbf{E_{ex}}$          & $\mathbf{\Phi_{ex}}$       & $\mathbf{d_{SiGeSn}}$        & $\mathbf{d_{GaAs}}$         & $\mathbf{\gamma}$      & $\mathbf{A_n}$           & $\mathbf{\Phi_{PR}}$     \\
 &  & (eV) & (eV) & (eV) & $\times 10^8$ & (rad) &  $\times 10^9$ & $\times 10^8$ & (meV) & (rad) & (nm) & (nm) & (eV$^{-1})$ & $\times 10^6$ & (rad) \\ \hline
A1         & 3.1 (0.4) & 0.94 (0.007) & 0.065 (0.003) & 0.284 (0.01) & 2.0 (0.5) & -1.6 (0.3) & 14.5 (4.0) & 8.6 (0.8) & 5 (2) & 0.4 (0.3) & 1811 (41) & n/a   & 1.2 (0.2) & 0.08 (0.01) & -1.3 (1.2) \\
A2         & 3.7 (0.3) & 1.001 (0.024) & 0.142 (0.004) & 0.273 (0.02) & 9.1 (0.6) & -1.2 (0.5) & 25.5 (14.0) & 3.9 (1.0) & 6 (2) & 1.2 (0.7) & 1746 (6) & n/a   & 1.8 (0.1) & 0.09 (0.01) & -0.1 (0.2) \\
A3         & 3.9 (0.0) & 1.014 (0.002) & 0.112 (0.006) & 0.294 (0.01) & 1.1 (0.2) & -1.9 (0.3) & 19.8 (5.0) & 9.3 (0.5) & 6 (2) & -2.3 (0.1) & 1828 (2) & n/a & 1.6 (0.1) & 0.09 (0.01) & -0.1 (0.2) \\
B1 - thin & 1.9 (0.5) & 0.866 (0.007) & 0.037 (0.002) & 0.291 (0.01) & 1.5 (0.4) & 1.1 (1.5) & 3.4 (2.0) & 3.9 (1.6) & 6 (3) & -0.6 (2.0) & 332 (11) & 503 (16) & 0.5 (0.4) & 2.18 (1.08) & -0.3 (1.5) \\
B1 - thick & 3.1 (0.4) & 0.866 (0.003) & 0.037 (0.002) & 0.29 (0.01) & 1.3 (0.3) & -2.5 (0.2) & 1.8 (2.0) & 4.3 (1.0) & 5 (1) & -0.4 (0.2) & 2073 (16) & 504 (4) & 0.4 (0.3) & 1.48 (0.48) & 2.3 (0.1)  \\
B2 - thin & 2.9 (0.4) & 0.911 (0.003) & 0.027 (0.001) & 0.301 (0.01) & 2.0 (0.2) & -0.8 (0.3) & 4.4 (2.0) & 7.9 (1.4) & 6 (2) & 1.7 (0.4) & 362 (8) & 502 (15) & 0.4 (0.3) & 2.28 (1.31) & -0.4 (1.3) \\
B2 - thick& 3.1 (0.4) & 0.907 (0.015) & 0.036 (0.002) & 0.292 (0.01) & 1.4 (0.1) & -0.3 (1.5) & 4.1 (2.0) & 4.6 (1.8) & 7 (2) & -0.1 (1.8) & 2017 (39) & 506 (8) & 0.7 (0.4) & 4.43 (1.85) & 1.5 (0.2)  \\
B3 - thin & 2.6 (0.6) & 0.953 (0.012) & 0.039 (0.001) & 0.292 (0.01) & 1.4 (0.4) & -1.4 (0.6) & 7.0 (1.0) & 3.3 (1.4) & 7 (2) & -0.0 (0.5) & 353 (7) & 515 (9) & 0.4 (0.3) & 2.45 (1.04) & 0.2 (1.4)  \\
B3 - thick & 3.6 (0.3) & 0.949 (0.008) & 0.035 (0.003) & 0.295 (0.01) & 1.3 (0.3) & 1.9 (0.6) & 3.2 (2.0) & 4.6 (2.0) & 6 (2) & -1.6 (1.7) & 2036 (49) & 497 (11) & 0.8 (0.5) & 3.67 (2.22) & 1.5 (0.2) 
\end{tabular}
\end{table}

\end{turnpage}

\clearpage

\section{Power-dependent photoluminescence measurements}

\begin{figure}[h]
	\centering
	\includegraphics[width=0.36\textwidth]{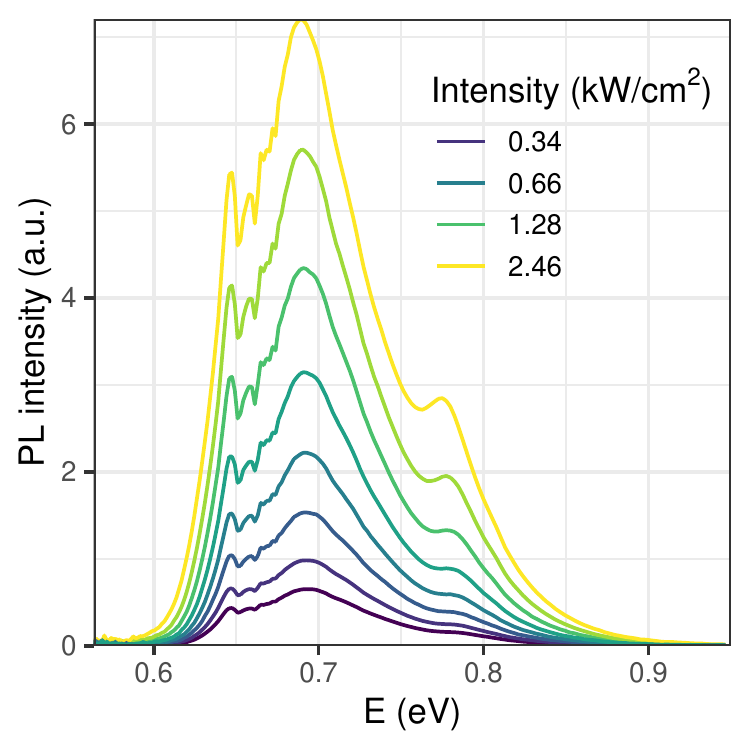}
	\caption{Measured power-dependent room temperature PL spectra of a reference Ge substrate sample. The incident laser intensity varies from 0.34 kW cm$^{-2}$ (dark purple) to 2.46 kW cm$^{-2}$ (yellow). In addition to the fundamental (indirect $\Gamma$-L) band gap emission at 0.69 eV, satellite features related to additional phonon-assisted emission processes are also visible.}
 	\label{fig:GePLpower}
\end{figure}

\begin{figure}[]
	\centering
	\includegraphics[width=\textwidth]{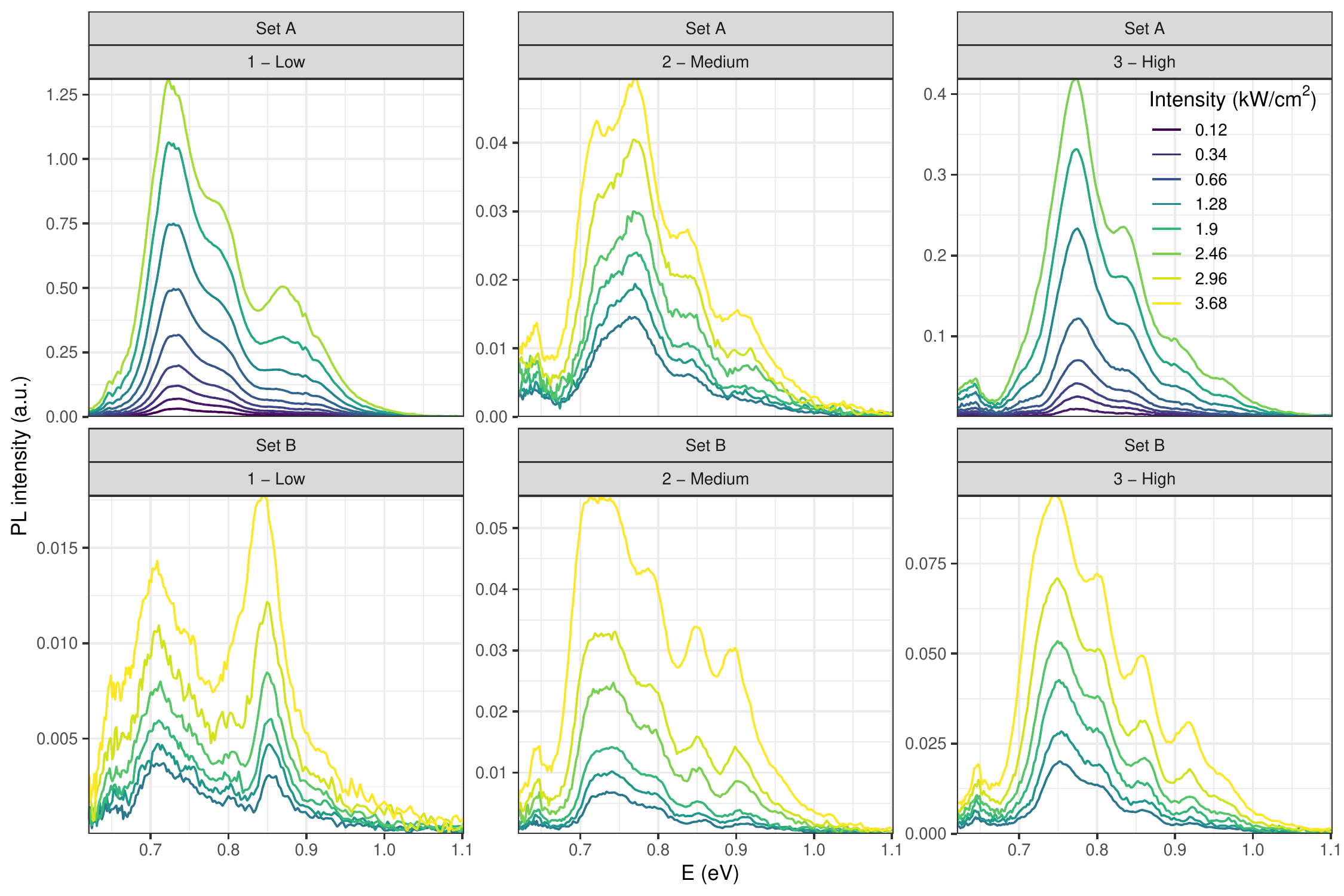}
	\caption{Measured power-dependent room temperature PL spectra of the  samples. Measurements for sample set B were performed on ``thick'' samples ($\approx$ 2 \si{\micro}m). Samples A1 and A3 had higher PL intensities at a given incident laser intensity than the other four samples, so PL measurements performed for incident laser intensities between 0.12 and 2.46 kW cm$^{-2}$. PL measurements for the other four samples were performed for incident laser intensities between 1.23 and 3.68 kW cm$^{-2}$.}
 	\label{fig:SiGeSnPLpower}
\end{figure}

\begin{figure}[]
	\centering
	\includegraphics[width=0.7\textwidth]{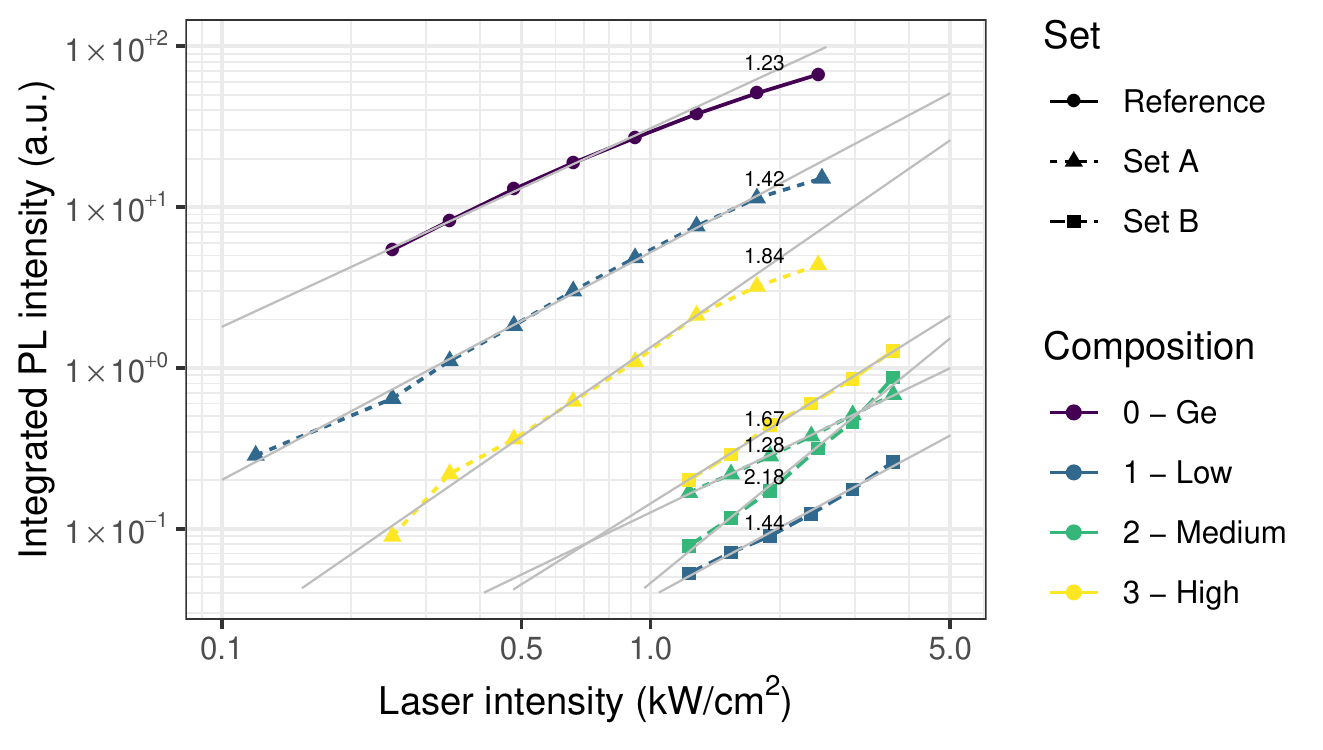}
	\caption{Variation of integrated room-temperature PL intensity over the whole measurement range (0.62 eV to 1.10 eV/1125 nm to 2000 nm) with incident laser intensity for each sample. The grey lines show fits to a power law (eq. \ref{eq:powerlaw}); the number by each line is the fitted value of $m$. For the higher PL intensity samples (Ge, A1 and A3), only data for powers up to and including 0.66 kW/cm$^2$ was included as the dependence above these powers becomes visibly sub-linear on the log-log plot.}
	\label{fig:powerintlog}
\end{figure}

Unnormalised power-dependent room-temperature PL measurements of the Ge and  samples are shown in Fig. \ref{fig:GePLpower} and \ref{fig:SiGeSnPLpower} respectively. The power dependence of the integrated PL intensity for each sample is shown in Fig. \ref{fig:powerintlog}.

\clearpage

\section{Local sample heating due to laser exposure}

To estimate the temperature increase due to the high-power laser (up to 3W total power, or 3.68 kW/cm$^2$ intensity, for some samples) used for the PL measurements, two methods were used: fitting the slope of the high-energy edge of the measured PL to find the carrier temperature, and solving the steady-state heat diffusion equation. 

\subsection{Obtaining the lattice temperature from PL data}
\begin{figure}[h]
	\centering
	\includegraphics[width=\textwidth]{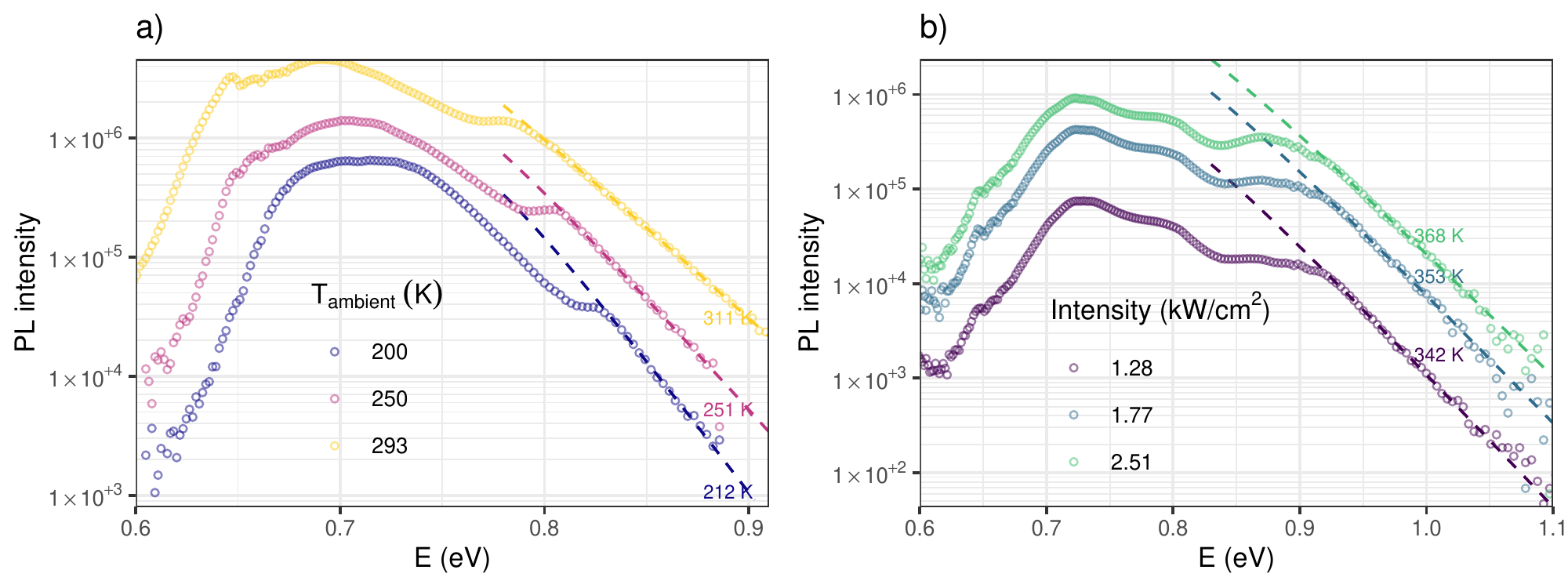}
	\caption{Fits of the high-energy PL edge to eq. \ref{eq:latticetemp}. a) Shows data for Ge, measured at varying temperatures (in a cryostat) with 1 W (1.28 kW/cm$^2$) of incident laser power (intensity).  b) Shows data for Si$_{x}$Ge$_{1-x-y}$Sn$_{y}$ sample A1, all measured at room temperature (not using a cryostat) but with laser power (intensity) between 1 W (1.28 kW/cm$^2$) and 3 W (2.51 kW/cm$^2$). In both plots, the labels next to the lines indicate the temperature obtained from fitting eq. \ref{eq:latticetemp}. The spectra are rigidly shifted along the y-axis for clarity.}
	\label{fig:PL_temp_fit}
\end{figure}

Following the method applied to Ge by Lieten et al. \cite{Lieten2012}, we fit the high-energy edge of the measured PL using:

\begin{equation}\label{eq:latticetemp}
    I_{PL} \propto \left(E-E_{0}\right)^{0.5} e^{-\left(E-E_{0}\right) / k T_c}
\end{equation}

The first term approximates the density of states above the bandgap, and the second term is a Boltzmann distribution where $T_c$ is the carrier temperature. Figure \ref{fig:PL_temp_fit}a shows the results of fitting this equation to data for a Ge reference sample measured at different temperatures using a laser power of 1 W. Below room temperature (200K and 250K), the measurements were carried out in a cryostat (i.e. the sample is actively being cooled), and the fitted $T_c$ values deviate by 12K and 1K from the nominal temperature. At room temperature (293K), no cryostat was used, and the $T_c$ fitted is 18K above room temperature. Fitting the high-energy PL edge in this manner thus gives reasonable results for the Ge reference sample. In Fig. \ref{fig:PL_temp_fit}b, PL of Si$_{x}$Ge$_{1-x-y}$Sn$_{y}$ sample A1 measured at room temperature (without a cryostat) at different lasers powers above 1W (1.28 kW/cm$^2$) is shown. Fitting the high-energy slope here gives $T_c$ values between 49K (1W laser power) and 60K (3W laser power) above room temperature, indicating there is quite significant heating of the samples due to laser exposure.

\subsection{Calculating the temperature due to laser exposure}

The steady-state (i.e. time-invariant) heat equation is:

\begin{equation}
    k~\nabla^2 T = \dot{q}
\end{equation}

Where $k$ is the conductivity of Ge (material constants for Ge are used throughout as an estimate for the  values), taken to be 0.58 W m$^{-1}$ K$^{-1}$, and $\dot{q}$ is the heat generated in the sample by the laser. In this case, we assume a Gaussian beam profile and Beer-Lambert absorption of the laser in the sample, and thus the problem can be reduced to two dimensions:

\begin{equation}\label{eq:heateq}
    k \left(\frac{\partial^2 T^2}{\partial x^2} + \frac{\partial^2 T^2}{\partial y^2} \right) = \frac{2P(1-R)\alpha}{\pi r_s^2}e^{-2(y-y_{inc})^2/r_s^2}e^{-\alpha x}
\end{equation}

Where $P$ is the total incident laser power, as measured, $R$ is the fraction of light reflected at the front surface (calculated to be 0.52 for Ge at 532 nm), $\alpha$ is the absorption coefficient of Ge at 532 nm ($5.5 \times 10^4$), and $r_s$ is the radius of the laser spot size where its intensity falls to $1/e^2$ of the maximum intensity. It was assumed that the material properties (conductivity and absorption coefficient) are constant. Integrating over $y$ and $x$ gives a total absorbed laser power of $P(1-R)$, as required.

To solve this partial differential equation (PDE), we use Firedrake \cite{Rathgeber2016}, which uses the Finite Element Method (FEM) to numerically solve PDEs (which can be expressed symbolically). The boundary conditions used assume that no heat can flow across the surfaces of the sample in contact with air (i.e. the front surface, where the laser is incident, and the sides) so that $\nabla u \cdot \overrightarrow{n} = 0$ where $\overrightarrow{n}$ is the unit vector pointing out of the surface. The rear surface allows heat flow and is maintained at the ambient temperature; the rear surface of the sample is in contact with a copper plate, which is much larger than the sample itself, and it is assumed that heat is conducted away from the sample efficiently by the copper. The thickness of the sample was fixed at 500 \si{\micro}{m}, and the width at 2.5 mm. The samples measured have variable thickness depending on the substrate used, and differ in width as samples for measurement were cleaved manually, but these values are representative for the whole sample set.

\begin{figure}[h]
	\centering
	\includegraphics[width=\textwidth]{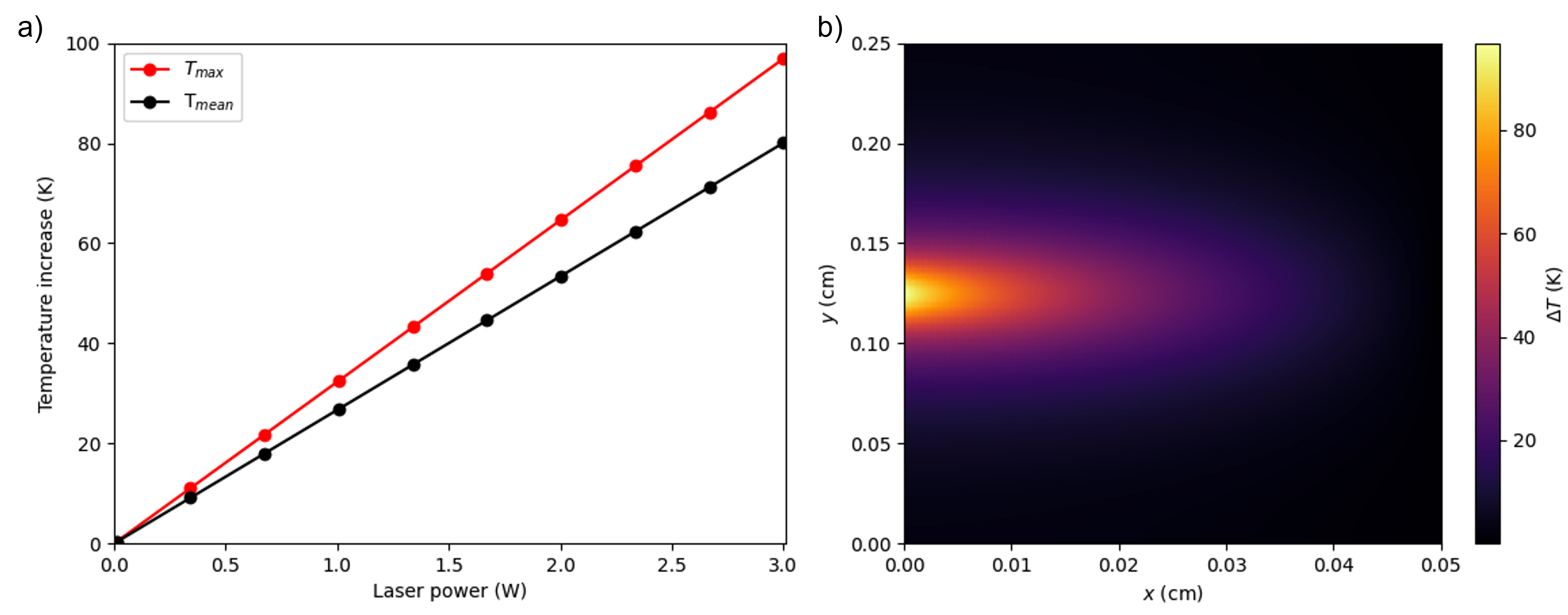}
	\caption{a) Maximum ($T_{max}$) and mean ($T_{mean}$) temperatures observed at the surface of the sample ($x = 0$) within the laser spot radius $r_s$ vs. incident laser power. b) Temperature increase (relative to room temperature) for the maximum incident laser power used (3 W). The laser is incident from the left at $x = 0$ cm, $y = 0.125$ cm}
	\label{fig:firedrake_sims}
\end{figure}

Results of the FEM calculations are shown in Fig. \ref{fig:firedrake_sims}. Fig. \ref{fig:firedrake_sims}a shows the maximum local temperature and the mean temperature within the radius of the laser spot found in at the sample surface at each incident laser power, while Fig. \ref{fig:firedrake_sims}b shows the spatial distribution of temperatures in the sample with 3W incident laser power. These results show that at high laser powers, there is expected to be significant heating of the sample in the area exposed to the laser, up to 100 K for the highest laser power used (3W). The maximum and mean temperatures observed at the sample surface scale linearly with the laser power. We expect these temperatures to be an overestimate as they ignore heat loss through radiation and convection.

The two methods for calculating the heat generated in the sample by the laser agree quite well; solving the heat equation gave a temperature increase above room temperature of around 25--30K at 1W laser power, and 50--60K at 2W. Looking at Fig. \ref{fig:PL_temp_fit}b, the carrier temperature fitted to eq. \ref{eq:latticetemp} for 1W and 2W incident laser power are 342K and 368K respectively. Taking room temperature to be 293K (20$^\circ$C), this gives temperature increases of 49K and 75K respectively.

\clearpage 

\section{Photoluminescence data fitting}

The Gaussian and Lorentzian peaks fitted to the PL data were expressed as:

\begin{equation}
I_G(E)=\frac{A}{\sigma \sqrt{2 \pi}} e^{-\frac{1}{2}\left(\frac{E-E_{i}}{\sigma}\right)^{2}}\label{eq:gaussian}
\end{equation}

\begin{equation}
I_{L}(E)=\frac{A \gamma}{\pi} \frac{1}{\left(E-E_{i}\right)^{2}+\gamma^{2}}\label{eq:lorentzian}
\end{equation} 

\begin{figure}[h]
	\centering
	\includegraphics[width=\textwidth]{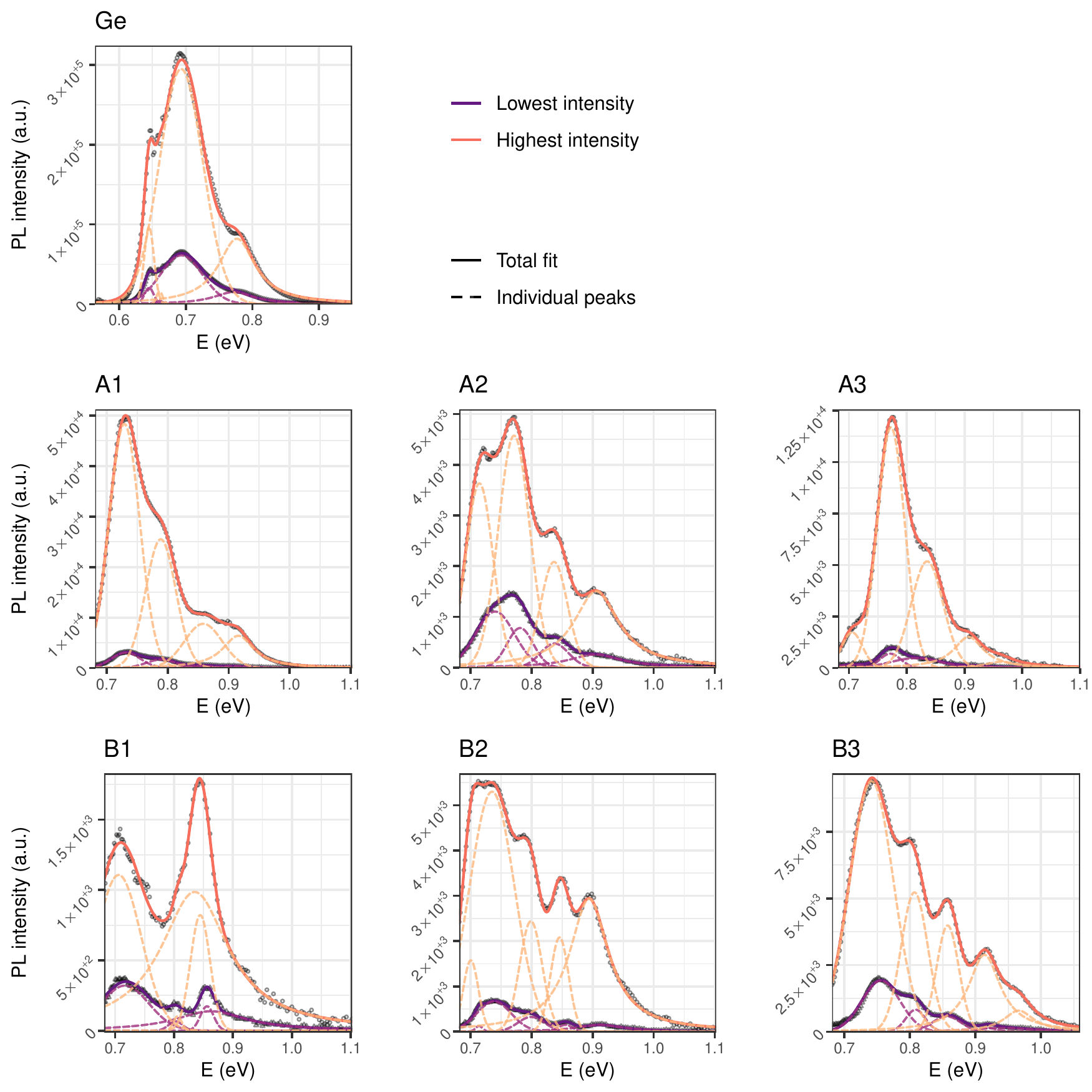}
	\caption{Fits to power-dependent room-temperature PL data using Gaussian and Lorentzian lineshapes. The lowest and highest power measured and the corresponding fits (total fit and individual peaks) are shown for each sample.}
	\label{fig:PLpeakfits}
\end{figure}

\begin{figure}[h]
	\centering
	\includegraphics[width=0.33\textwidth]{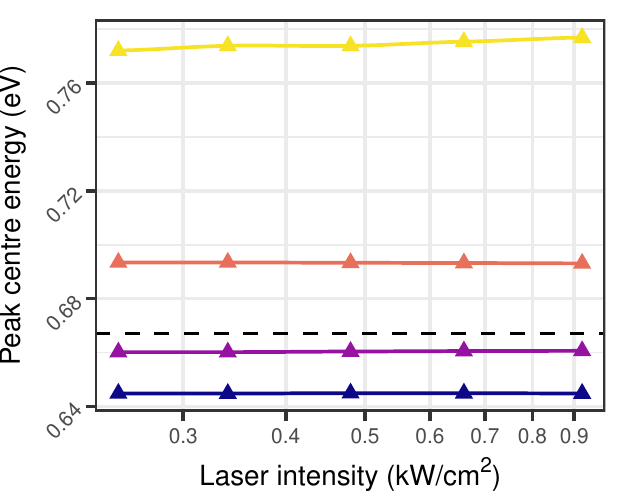}
	\caption{Dependence of fitted peak centre energy on incident laser intensity for Ge.}
	\label{fig:Ge_E}
\end{figure}

\begin{figure}[h]
	\centering
	\includegraphics[width=\textwidth]{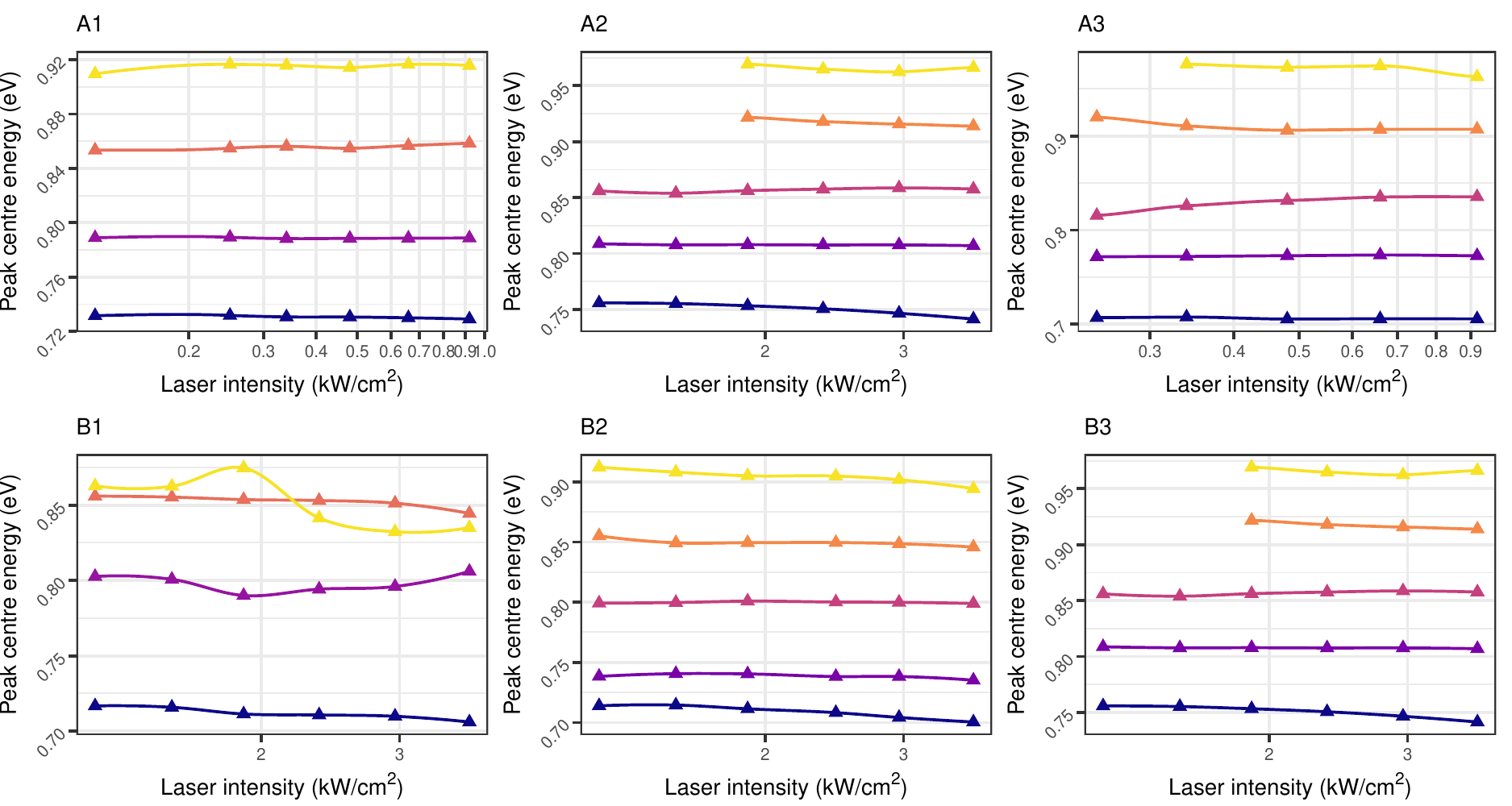}
	\caption{Dependence of fitted peak centre energy on incident laser intensity for the Si$_{x}$Ge$_{1-x-y}$Sn$_{y}$ samples.}
	\label{fig:SiGeSn_E}
\end{figure}

The parameters fitted in each case were the centre energy of the peak ($E_i$), the broadening (standard deviation $\sigma$ for the Gaussian distributions and scale parameter $\gamma$ for the Lorentzian distribution), and the strength $A$, which corresponds to the area under the peak. Figure \ref{fig:PLpeakfits} shows the resulting fits at the highest and lowest incident laser power used to measure each sample. In each case, it was possible to fit the same number and type of peak consistently across data for different incident powers, although for sample A2 it was not possible to fit the two lowest-energy peaks at low power. Figures \ref{fig:Ge_E} and \ref{fig:SiGeSn_E} show how the centre energies fitted for each peak change with the laser power used; this shows that despite the expected increase in temperature due to the laser, the centre energies fitted are quite constant, although there is some evidence of a general redshift with increasing power for the samples measured at the highest powers (A2, B1, B2 and B3). To avoid issues due to sample heating and band-filling as much as possible, data for the lowest power possible was used to estimate the band gap as presented in Table II of the main manuscript.

The data in Fig. \ref{fig:SiGeSnPLpower} shows that a peak at 0.645 eV/1922 nm is present for all Si$_{x}$Ge$_{1-x-y}$Sn$_{y}$ samples, independent of the laser power used to excite the samples. It is not clearly visible in all the plots in \ref{fig:SiGeSnPLpower} due to the varying intensity of the PL. Since this peak was present for all measurements, it was attributed to some background signal (note that it was not possible to see this peak in the measurements for Ge – there is significant PL intensity from the Ge samples at 0.645 eV, and the PL signal is around 2 orders of magnitude higher than for the samples where this constant-energy peak is apparent, see Figs S2 and S3). This peak did not have the expected power dependence for PL for a bulk material as observed for the other peaks; for instance, sample B3 has around 5 times the PL intensity of sample B1 at a given power (Fig. S3), yet the peaks around 0.645 eV are only around twice as intense, indicating that their source is some external source of background signal with constant intensity (with the remaining difference of intensity being due to the low-energy tail of the PL signal). For this reason, only data above 0.68 eV was used in fitting the PL signal of the samples with relatively low PL intensity (A2, B1, B2 and B3). 

Figure \ref{fig:powerintlog} shows the variation of the integrated PL intensity (area under the curve) for all samples, integrated over the whole measurement range (0.62 eV to 1.10 eV/1125 nm to 2000 nm). The room temperature power dependence of the integrated PL intensity of all samples follow a power law of the form:

\begin{equation}\label{eq:powerlaw}
    I_{PL} \propto I_{laser}^m
\end{equation}

% not sure if the peak power is relevant. Not a metric people normally use.
% should include fact that signal is corrected for detector response.

where $I_{PL}$ is the integrated intensity of the PL signal and $I_{laser}$ is the incident laser intensity (the peak intensity follows a similar power law behaviour for all samples) and $m$ is a positive constant. The samples which showed higher PL intensities (Ge, A1 and A3) become sub-linear on a log-log plot at powers above $\approx$ 1 kW/cm$^2$, indicating that at higher powers there may be additional effects such as band-filling which affect the PL intensity. The power law fits shown in Fig. \ref{fig:powerintlog} are fitted to data for all the incident intensities for the lower-intensity samples (A2, B1, B2 and B3) and up to and including the measurements at 0.66 kW/cm$^2$ for the higher-intensity samples (Ge, A1 and A3).

\begin{table}[h]
\small
\caption{\label{tab:exponents}Power law exponents fitted to the peak and integrated intensity of PL data for Ge and  samples.}
\begin{tabular}{lll}
Sample & $m_{int}$ & $m_{peak}$ \\ \hline
Ge     & 1.23      & 1.20       \\
A1     & 1.42      & 1.39       \\
A2     & 1.28      & 1.11       \\
A3     & 1.83      & 1.84       \\
B1     & 1.43      & 1.53       \\
B2     & 2.17      & 1.88       \\
B3     & 1.67      & 1.38      
\end{tabular}
\end{table}

\begin{table}[h]
\caption{\label{tab:Gesummary}Summary of some parameters from this and previous work on Ge photoluminescence.}
\begin{tabular}{llll}
Reference & Power law exponent & Power densities & Temperature (K)\\ \hline
Lieten et al. \cite{Lieten2012}    & 2.02 (indirect) & 0.008--0.128 kW/cm$^2$ & 9\\
Klingenstein \& Schweizer \cite{Klingenstein1978} & 1 (indirect), 1.6 (direct) & 0.8--6.7 kW/cm$^2$ & 2 \\
van Driel et al. \cite{VanDriel1976} & not reported & 0.05--10$^5$ kW/cm$^2$ & 174, 295 \\
This work - 295 K & 1.23 (indirect) & 0.12--3.68 kW/cm$^2$ & 295  \\
This work - 30 K & 1.68 (indirect) & 0.01--0.25 kW/cm$^2$ & 30
\end{tabular}
\end{table}

While PL from all the samples follows a power law as expected, the variation of the exponent $m$ between samples does not seem to correlate to an obvious parameter such as peak energy, composition or Si:Sn ratio, and lies between 1 and 2 for all the samples. The Ge substrate has an exponent of 1.23, which differs from values previously reported in the literature as summarized in Table \ref{tab:Gesummary}; Lieten et al. \cite{Lieten2012} report a power law exponent of 2.02, relating to PL from the indirect transition, while Klingenstein \& Schweizer \cite{Klingenstein1978} report an exponent of 1 for the indirect transition and 1.6 for the direct transition, with the latter measured at significantly higher laser intensities by around two orders of magnitude (Table \ref{tab:Gesummary}). Both of these sets of measurements were taken at low temperature (9K and 2K respectively). Measurements of the Ge sample at 30 K give an exponent of 1.66, higher than the room-temperature value of 1.23. The laser intensity used here for the low-temperature measurements is between those used by Lieten et al. and Klingenstein \& Schweizer, and an intermediate value of the power law exponent is obtained. Considering the power-dependent data for Ge at room temperature shown here, and the exponents for the indirect transition obtained by Lieten at low laser intensity and by Klingenstein at high power, it appears that higher incident laser power decreases the exponent $m$ while higher temperatures lead to higher $m$. This indicates that the behaviour of the Ge PL is quite consistent with that previously observed once variations of the exponent $m$ with temperature and incident laser intensity are taken into account.

Schmidt et al. \cite{Schmidt1992} formulated a general model to describe the dependence of the intensity of the near-bandgap luminescence on the incident laser power by considering rate equations for carrier excitation and different recombination processes, including free and bound excitons and recombination involving donors and acceptors, and non-radiative processes; this model indicates that a value of $m$ between 1 and 2 is consistent with interband radiative recombination, rather than transitions involving bound states (such as would occur with defects), indicating that the PL emitted by the Ge and  samples at room temperature is consistent with radiative recombination in the bulk.

\clearpage

%%%%%%%%%%%%%%%%%%%%%%%%%%%%%%%%%%%%%%%%%%%%%
%%%% Section 2: Theoretical calculations %%%%
%%%%%%%%%%%%%%%%%%%%%%%%%%%%%%%%%%%%%%%%%%%%%

\section{Character of Si$_{x}$Ge$_{1-x-y}$Sn$_{y}$ alloy valence band states from tight-binding supercell calculations}\label{sec:theory}

In Fig.~4 of the main text we presented the calculated indirect (Ge L$_{6c}$ and X$_{5c}$) and direct (Ge $\Gamma_{7c}$) character of Si$_{x}$Ge$_{1-x-y}$Sn$_{y}$ alloy states lying close in energy to the conduction band (CB) edge. These calculations demonstrated the presence of strong alloy-induced band mixing in the CB, leading to significant energetic broadening of the associated L-, X- and $\Gamma$-point CB Bloch character. Here, we present the corresponding calculated character of alloy states lying close in energy to the valence band (VB) edge, demonstrating that the VB structure is much less strongly perturbed.

\begin{figure*}[h!]
	\includegraphics[width=1.00\textwidth]{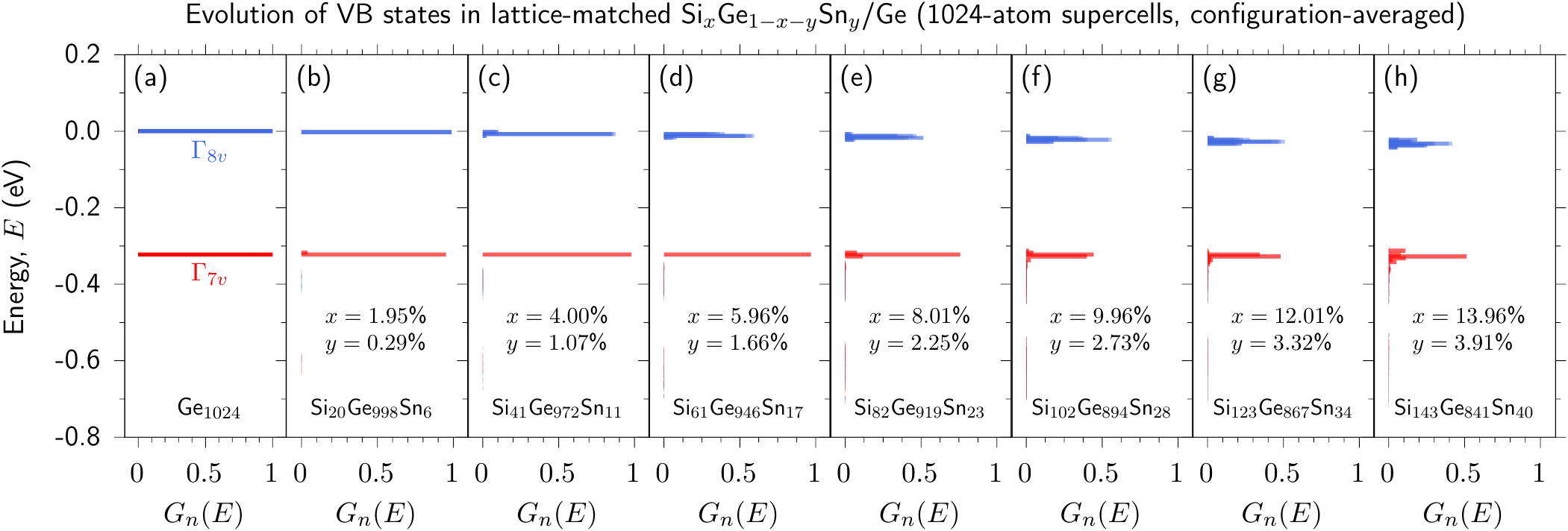}
	\caption{Calculated evolution of the configuration-averaged Ge $\Gamma_{8v}$ (blue) and $\Gamma_{7v}$ (red) character of the VB states in disordered, 1024-atom Si$_{x}$Ge$_{1-x-y}$Sn$_{y}$ alloy supercells lattice-matched to Ge ($x$:$y = 3.71$:1). Panel (a) illustrates for reference that the HH/LH (blue) and SO (red) VB edge states in pure Ge respectively possess 100\% Ge $\Gamma_{8v}$ and $\Gamma_{7v}$ character. Substitutional incorporation of Si and Sn -- panels (b)--(h) -- produces minimal changes in the band structure close in energy to the VB edge. The reduction in symmetry associated with short-range alloy disorder results in inhomogeneous energetic broadening of the associated VB edge features, but much more weakly than for the alloy CB edge states. The zero of energy is taken to lie at the Ge VB edge.}
	\label{fig:vb_character}
\end{figure*}

For the VB we take as our Ge host matrix eigenstates $\vert n^{(0)} \rangle$ the $\Gamma_{8v}$ heavy-hole (HH) and light-hole (LH) states, and the $\Gamma_{7v}$ spin-orbit split-off (SO) states. Using these states in conjunction with Eq.~(2) of the main text, we compute the Ge HH/LH and SO character of the alloy VB states $\vert m \rangle$. We note that these VB calculations were performed using the same alloy supercells, and hence alloy eigenstates $\vert m \rangle$, and energy binning employed for the calculation of the CB character of the main text. The results of our VB calculations are summarised in Fig.~\ref{fig:vb_character}, which displays the calculated Ge $\Gamma_{8v}$ (HH/LH; blue) and $\Gamma_{7v}$ (SO; red) character. As in Fig.~4 of the main text, the results presented in Fig.~\ref{fig:vb_character} have been configuration-averaged.

Beginning in panel (a), we note for reference that the HH/LH and SO states in a pure Ge supercell respectively possess 100\% $\Gamma_{8v}$ and $\Gamma_{7v}$ character (i.e.~$G_{n} = 1$). As the Si and Sn composition increases in panels (b) -- (h) we note (i) a small reduction in energy of the HH/LH and SO VB edge energies, and (ii) an energetic broadening of the Ge $\Gamma_{8v}$ and $\Gamma_{7v}$ character. Here, (i) is associatfed with minimal change in the VB spin-orbit splitting energy $\Delta_{0}$, while (ii) is significantly less pronounced than for the Ge L$_{6c}$, X$_{5c}$ and $\Gamma_{7c}$ character of the alloy CB states (cf.~Fig.~4 of the main text). Indeed, we find that the energetic broadening of the Ge $\Gamma_{8v}$ and $\Gamma_{7v}$ character of the alloy VB states is minimal, with almost all of the calculated Ge HH/LH or SO character lying within a small energy range; note, e.g., the difference in scales on the abscissae of Fig.~\ref{fig:vb_character} (below) vs.~Fig.~4 (main text). This indicates that the observed limited energetic broadening is driven primarily by the weak perturbation associated with the reduction in translational symmetry in the disordered alloy, rather than by a strong perturbation associated with alloy-induced hybridisation (as dominates the band structure close in energy to the CB edge).

As such, we conclude that the VB edge states in lattice-matched Si$_{x}$Ge$_{1-x-y}$Sn$_{y}$/Ge alloys retain primarily Ge-like character, being minimally perturbed from those of the Ge host matrix semiconductor at the compositions of interest for photovoltaics (i.e.~when the direct band gap is $\sim 1$ eV). From a theoretical perspective, our results then suggest overall that while the virtual crystal approximation (VCA) can likely be applied to model the Si$_{x}$Ge$_{1-x-y}$Sn$_{y}$ VB structure with reasonable accuracy, beyond-VCA approaches -- i.e.~direct atomistic calculations or appropriately modified continuum model Hamiltonians -- are required in order to quantitatively describe the nature and evolution of the CB states, and hence the alloy band gap and its optical properties.

\clearpage

%%%%%%%%%%%%%%%%%%%%
%%%% References %%%%
%%%%%%%%%%%%%%%%%%%%

\bibliographystyle{apsrev}               % Using the Physical Review style for references
\bibliography{supplementary} % The bibliography is contained in the file: SiGeSn_supplementary.bib